\documentclass{article}

    \usepackage{graphicx} 
    \usepackage{adjustbox} 
    \usepackage{color} 
    \usepackage{enumerate} 
    \usepackage{geometry} 
    \usepackage{amsmath} 
    \usepackage{amssymb} 
    \usepackage{eurosym} 
    \usepackage[mathletters]{ucs} 
    \usepackage[utf8x]{inputenc} 
    \usepackage{fancyvrb} 
    \usepackage{grffile} 
    \usepackage{hyperref}
    \usepackage{longtable} 
    \usepackage{booktabs}  

    \definecolor{orange}{cmyk}{0,0.4,0.8,0.2}
    \definecolor{darkorange}{rgb}{.71,0.21,0.01}
    \definecolor{darkgreen}{rgb}{.12,.54,.11}
    \definecolor{myteal}{rgb}{.26, .44, .56}
    \definecolor{gray}{gray}{0.45}
    \definecolor{lightgray}{gray}{.95}
    \definecolor{mediumgray}{gray}{.8}
    \definecolor{inputbackground}{rgb}{.95, .95, .85}
    \definecolor{outputbackground}{rgb}{.95, .95, .95}
    \definecolor{traceback}{rgb}{1, .95, .95}
    \definecolor{red}{rgb}{.6,0,0}
    \definecolor{green}{rgb}{0,.65,0}
    \definecolor{brown}{rgb}{0.6,0.6,0}
    \definecolor{blue}{rgb}{0,.145,.698}
    \definecolor{purple}{rgb}{.698,.145,.698}
    \definecolor{cyan}{rgb}{0,.698,.698}
    \definecolor{lightgray}{gray}{0.5}

    \definecolor{darkgray}{gray}{0.25}
    \definecolor{lightred}{rgb}{1.0,0.39,0.28}
    \definecolor{lightgreen}{rgb}{0.48,0.99,0.0}
    \definecolor{lightblue}{rgb}{0.53,0.81,0.92}
    \definecolor{lightpurple}{rgb}{0.87,0.63,0.87}
    \definecolor{lightcyan}{rgb}{0.5,1.0,0.83}

    \DefineVerbatimEnvironment{Highlighting}{Verbatim}{commandchars=\\\{\}}
    \newenvironment{Shaded}{}{}
    \newcommand{\KeywordTok}[1]{\textcolor[rgb]{0.00,0.44,0.13}{\textbf{{#1}}}}
    
    \newcommand{\DecValTok}[1]{\textcolor[rgb]{0.25,0.63,0.44}{{#1}}}
    
    \newcommand{\FloatTok}[1]{\textcolor[rgb]{0.25,0.63,0.44}{{#1}}}
    \newcommand{\CharTok}[1]{\textcolor[rgb]{0.25,0.44,0.63}{{#1}}}
    \newcommand{\StringTok}[1]{\textcolor[rgb]{0.25,0.44,0.63}{{#1}}}
    \newcommand{\CommentTok}[1]{\textcolor[rgb]{0.38,0.63,0.69}{\textit{{#1}}}}

    \newcommand{\NormalTok}[1]{{#1}}

    \def\lt{<}
    \title{UQ_Handbook_Chapter}

\makeatletter
\def\PY@reset{\let\PY@it=\relax \let\PY@bf=\relax%
    \let\PY@ul=\relax \let\PY@tc=\relax%
    \let\PY@bc=\relax \let\PY@ff=\relax}
\def\PY@tok#1{\csname PY@tok@#1\endcsname}
\def\PY@toks#1+{\ifx\relax#1\empty\else%
    \PY@tok{#1}\expandafter\PY@toks\fi}
\def\PY@do#1{\PY@bc{\PY@tc{\PY@ul{%
    \PY@it{\PY@bf{\PY@ff{#1}}}}}}}
\def\PY#1#2{\PY@reset\PY@toks#1+\relax+\PY@do{#2}}

\expandafter\def\csname PY@tok@gd\endcsname{\def\PY@tc##1{\textcolor[rgb]{0.63,0.00,0.00}{##1}}}
\expandafter\def\csname PY@tok@gu\endcsname{\let\PY@bf=\textbf\def\PY@tc##1{\textcolor[rgb]{0.50,0.00,0.50}{##1}}}
\expandafter\def\csname PY@tok@gt\endcsname{\def\PY@tc##1{\textcolor[rgb]{0.00,0.27,0.87}{##1}}}
\expandafter\def\csname PY@tok@gs\endcsname{\let\PY@bf=\textbf}
\expandafter\def\csname PY@tok@gr\endcsname{\def\PY@tc##1{\textcolor[rgb]{1.00,0.00,0.00}{##1}}}
\expandafter\def\csname PY@tok@cm\endcsname{\let\PY@it=\textit\def\PY@tc##1{\textcolor[rgb]{0.25,0.50,0.50}{##1}}}
\expandafter\def\csname PY@tok@vg\endcsname{\def\PY@tc##1{\textcolor[rgb]{0.10,0.09,0.49}{##1}}}
\expandafter\def\csname PY@tok@m\endcsname{\def\PY@tc##1{\textcolor[rgb]{0.40,0.40,0.40}{##1}}}
\expandafter\def\csname PY@tok@mh\endcsname{\def\PY@tc##1{\textcolor[rgb]{0.40,0.40,0.40}{##1}}}
\expandafter\def\csname PY@tok@go\endcsname{\def\PY@tc##1{\textcolor[rgb]{0.53,0.53,0.53}{##1}}}
\expandafter\def\csname PY@tok@ge\endcsname{\let\PY@it=\textit}
\expandafter\def\csname PY@tok@vc\endcsname{\def\PY@tc##1{\textcolor[rgb]{0.10,0.09,0.49}{##1}}}
\expandafter\def\csname PY@tok@il\endcsname{\def\PY@tc##1{\textcolor[rgb]{0.40,0.40,0.40}{##1}}}
\expandafter\def\csname PY@tok@cs\endcsname{\let\PY@it=\textit\def\PY@tc##1{\textcolor[rgb]{0.25,0.50,0.50}{##1}}}
\expandafter\def\csname PY@tok@cp\endcsname{\def\PY@tc##1{\textcolor[rgb]{0.74,0.48,0.00}{##1}}}
\expandafter\def\csname PY@tok@gi\endcsname{\def\PY@tc##1{\textcolor[rgb]{0.00,0.63,0.00}{##1}}}
\expandafter\def\csname PY@tok@gh\endcsname{\let\PY@bf=\textbf\def\PY@tc##1{\textcolor[rgb]{0.00,0.00,0.50}{##1}}}
\expandafter\def\csname PY@tok@ni\endcsname{\let\PY@bf=\textbf\def\PY@tc##1{\textcolor[rgb]{0.60,0.60,0.60}{##1}}}
\expandafter\def\csname PY@tok@nl\endcsname{\def\PY@tc##1{\textcolor[rgb]{0.63,0.63,0.00}{##1}}}
\expandafter\def\csname PY@tok@nn\endcsname{\let\PY@bf=\textbf\def\PY@tc##1{\textcolor[rgb]{0.00,0.00,1.00}{##1}}}
\expandafter\def\csname PY@tok@no\endcsname{\def\PY@tc##1{\textcolor[rgb]{0.53,0.00,0.00}{##1}}}
\expandafter\def\csname PY@tok@na\endcsname{\def\PY@tc##1{\textcolor[rgb]{0.49,0.56,0.16}{##1}}}
\expandafter\def\csname PY@tok@nb\endcsname{\def\PY@tc##1{\textcolor[rgb]{0.00,0.50,0.00}{##1}}}
\expandafter\def\csname PY@tok@nc\endcsname{\let\PY@bf=\textbf\def\PY@tc##1{\textcolor[rgb]{0.00,0.00,1.00}{##1}}}
\expandafter\def\csname PY@tok@nd\endcsname{\def\PY@tc##1{\textcolor[rgb]{0.67,0.13,1.00}{##1}}}
\expandafter\def\csname PY@tok@ne\endcsname{\let\PY@bf=\textbf\def\PY@tc##1{\textcolor[rgb]{0.82,0.25,0.23}{##1}}}
\expandafter\def\csname PY@tok@nf\endcsname{\def\PY@tc##1{\textcolor[rgb]{0.00,0.00,1.00}{##1}}}
\expandafter\def\csname PY@tok@si\endcsname{\let\PY@bf=\textbf\def\PY@tc##1{\textcolor[rgb]{0.73,0.40,0.53}{##1}}}
\expandafter\def\csname PY@tok@s2\endcsname{\def\PY@tc##1{\textcolor[rgb]{0.73,0.13,0.13}{##1}}}
\expandafter\def\csname PY@tok@vi\endcsname{\def\PY@tc##1{\textcolor[rgb]{0.10,0.09,0.49}{##1}}}
\expandafter\def\csname PY@tok@nt\endcsname{\let\PY@bf=\textbf\def\PY@tc##1{\textcolor[rgb]{0.00,0.50,0.00}{##1}}}
\expandafter\def\csname PY@tok@nv\endcsname{\def\PY@tc##1{\textcolor[rgb]{0.10,0.09,0.49}{##1}}}
\expandafter\def\csname PY@tok@s1\endcsname{\def\PY@tc##1{\textcolor[rgb]{0.73,0.13,0.13}{##1}}}
\expandafter\def\csname PY@tok@sh\endcsname{\def\PY@tc##1{\textcolor[rgb]{0.73,0.13,0.13}{##1}}}
\expandafter\def\csname PY@tok@sc\endcsname{\def\PY@tc##1{\textcolor[rgb]{0.73,0.13,0.13}{##1}}}
\expandafter\def\csname PY@tok@sx\endcsname{\def\PY@tc##1{\textcolor[rgb]{0.00,0.50,0.00}{##1}}}
\expandafter\def\csname PY@tok@bp\endcsname{\def\PY@tc##1{\textcolor[rgb]{0.00,0.50,0.00}{##1}}}
\expandafter\def\csname PY@tok@c1\endcsname{\let\PY@it=\textit\def\PY@tc##1{\textcolor[rgb]{0.25,0.50,0.50}{##1}}}
\expandafter\def\csname PY@tok@kc\endcsname{\let\PY@bf=\textbf\def\PY@tc##1{\textcolor[rgb]{0.00,0.50,0.00}{##1}}}
\expandafter\def\csname PY@tok@c\endcsname{\let\PY@it=\textit\def\PY@tc##1{\textcolor[rgb]{0.25,0.50,0.50}{##1}}}
\expandafter\def\csname PY@tok@mf\endcsname{\def\PY@tc##1{\textcolor[rgb]{0.40,0.40,0.40}{##1}}}
\expandafter\def\csname PY@tok@err\endcsname{\def\PY@bc##1{\setlength{\fboxsep}{0pt}\fcolorbox[rgb]{1.00,0.00,0.00}{1,1,1}{\strut ##1}}}
\expandafter\def\csname PY@tok@kd\endcsname{\let\PY@bf=\textbf\def\PY@tc##1{\textcolor[rgb]{0.00,0.50,0.00}{##1}}}
\expandafter\def\csname PY@tok@ss\endcsname{\def\PY@tc##1{\textcolor[rgb]{0.10,0.09,0.49}{##1}}}
\expandafter\def\csname PY@tok@sr\endcsname{\def\PY@tc##1{\textcolor[rgb]{0.73,0.40,0.53}{##1}}}
\expandafter\def\csname PY@tok@mo\endcsname{\def\PY@tc##1{\textcolor[rgb]{0.40,0.40,0.40}{##1}}}
\expandafter\def\csname PY@tok@kn\endcsname{\let\PY@bf=\textbf\def\PY@tc##1{\textcolor[rgb]{0.00,0.50,0.00}{##1}}}
\expandafter\def\csname PY@tok@mi\endcsname{\def\PY@tc##1{\textcolor[rgb]{0.40,0.40,0.40}{##1}}}
\expandafter\def\csname PY@tok@gp\endcsname{\let\PY@bf=\textbf\def\PY@tc##1{\textcolor[rgb]{0.00,0.00,0.50}{##1}}}
\expandafter\def\csname PY@tok@o\endcsname{\def\PY@tc##1{\textcolor[rgb]{0.40,0.40,0.40}{##1}}}
\expandafter\def\csname PY@tok@kr\endcsname{\let\PY@bf=\textbf\def\PY@tc##1{\textcolor[rgb]{0.00,0.50,0.00}{##1}}}
\expandafter\def\csname PY@tok@s\endcsname{\def\PY@tc##1{\textcolor[rgb]{0.73,0.13,0.13}{##1}}}
\expandafter\def\csname PY@tok@kp\endcsname{\def\PY@tc##1{\textcolor[rgb]{0.00,0.50,0.00}{##1}}}
\expandafter\def\csname PY@tok@w\endcsname{\def\PY@tc##1{\textcolor[rgb]{0.73,0.73,0.73}{##1}}}
\expandafter\def\csname PY@tok@kt\endcsname{\def\PY@tc##1{\textcolor[rgb]{0.69,0.00,0.25}{##1}}}
\expandafter\def\csname PY@tok@ow\endcsname{\let\PY@bf=\textbf\def\PY@tc##1{\textcolor[rgb]{0.67,0.13,1.00}{##1}}}
\expandafter\def\csname PY@tok@sb\endcsname{\def\PY@tc##1{\textcolor[rgb]{0.73,0.13,0.13}{##1}}}
\expandafter\def\csname PY@tok@k\endcsname{\let\PY@bf=\textbf\def\PY@tc##1{\textcolor[rgb]{0.00,0.50,0.00}{##1}}}
\expandafter\def\csname PY@tok@se\endcsname{\let\PY@bf=\textbf\def\PY@tc##1{\textcolor[rgb]{0.73,0.40,0.13}{##1}}}
\expandafter\def\csname PY@tok@sd\endcsname{\let\PY@it=\textit\def\PY@tc##1{\textcolor[rgb]{0.73,0.13,0.13}{##1}}}

\def\PYZus{\char`\_}
\def\PYZob{\char`\{}
\def\PYZcb{\char`\}}

\def\PYZgt{\char`\>}
\def\PYZsh{\char`\#}
\def\PYZpc{\char`\%}

\def\PYZhy{\char`\-}
\def\PYZsq{\char`\'}
\def\PYZdq{\char`\"}
\def\PYZti{\char`\~}

\makeatother

    \definecolor{incolor}{rgb}{0.0, 0.0, 0.5}
    \definecolor{outcolor}{rgb}{0.545, 0.0, 0.0}

    \hypersetup{
      breaklinks=true,  
      colorlinks=true,
      urlcolor=blue,
      linkcolor=darkorange,
      citecolor=darkgreen,
      }

    \title{Probabilistic Programming in Python using PyMC}\label{probabilistic-programming-in-python-using-pymc}

\author{John Salvatier, Thomas V. Wiecki, Christopher Fonnesbeck}

\begin{document}

    \maketitle

\section{Introduction}\label{introduction}

Probabilistic programming (PP) allows flexible specification of Bayesian
statistical models in code. PyMC3 is a new, open-source PP framework
with an intutive and readable, yet powerful, syntax that is close to the
natural syntax statisticians use to describe models. It features
next-generation Markov chain Monte Carlo (MCMC) sampling algorithms such
as the No-U-Turn Sampler (NUTS; Hoffman, 2014), a self-tuning variant of
Hamiltonian Monte Carlo (HMC; Duane, 1987). This class of samplers works
well on high dimensional and complex posterior distributions and allows
many complex models to be fit without specialized knowledge about
fitting algorithms. HMC and NUTS take advantage of gradient information
from the likelihood to achieve much faster convergence than traditional
sampling methods, especially for larger models. NUTS also has several
self-tuning strategies for adaptively setting the tunable parameters of
Hamiltonian Monte Carlo, whicstatisticalh means you usually don't need
to have specialized knowledge about how the algorithms work. PyMC3, Stan
(Stan Development Team, 2014), and the LaplacesDemon package for R are
currently the only PP packages to offer HMC.

Probabilistic programming in Python confers a number of advantages
including multi-platform compatibility, an expressive yet clean and
readable syntax, easy integration with other scientific libraries, and
extensibility via C, C++, Fortran or Cython. These features make it
relatively straightforward to write and use custom statistical
distributions, samplers and transformation functions, as required by
Bayesian analysis.

While most of PyMC3's user-facing features are written in pure Python,
it leverages Theano (Bergstra et al., 2010) to transparently transcode
models to C and compile them to machine code, thereby boosting
performance. Theano is a library that allows expressions to be defined
using generalized vector data structures called \emph{tensors}, which
are tightly integrated with the popular NumPy \texttt{ndarray} data
structure, and similarly allow for broadcasting and advanced indexing,
just as NumPy arrays do. Theano also automatically optimizes the
likelihood's computational graph for speed and provides simple GPU
integration.

Here, we present a primer on the use of PyMC3 for solving general
Bayesian statistical inference and prediction problems. We will first
see the basics of how to use PyMC3, motivated by a simple example:
installation, data creation, model definition, model fitting and
posterior analysis. Then we will cover two case studies and use them to
show how to define and fit more sophisticated models. Finally we will
show how to extend PyMC3 and discuss other useful features: the
Generalized Linear Models subpackage, custom distributions, custom
transformations and alternative storage backends.

    \section{Installation}\label{installation}

Running PyMC3 requires a working Python interpreter, either version 2.7
(or more recent) or 3.4 (or more recent); we recommend that new users
install version 3.4. A complete Python installation for Mac OSX, Linux
and Windows can most easily be obtained by downloading and installing
the free
\href{https://store.continuum.io/cshop/anaconda/}{\texttt{Anaconda Python Distribution}}
by ContinuumIO.

\texttt{PyMC3} can be installed using \texttt{pip}
(https://pip.pypa.io/en/latest/installing.html):

\begin{verbatim}
pip install git+https://github.com/pymc-devs/pymc3
\end{verbatim}

PyMC3 depends on several third-party Python packages which will be
automatically installed when installing via pip. The four required
dependencies are: \texttt{Theano}, \texttt{NumPy}, \texttt{SciPy}, and
\texttt{Matplotlib}.

To take full advantage of PyMC3, the optional dependencies
\texttt{Pandas} and \texttt{Patsy} should also be installed. These are
\emph{not} automatically installed, but can be installed by:

\begin{verbatim}
pip install patsy pandas
\end{verbatim}

The source code for PyMC3 is hosted on GitHub at
https://github.com/pymc-devs/pymc3 and is distributed under the liberal
\href{https://github.com/pymc-devs/pymc3/blob/master/LICENSE}{Apache
License 2.0}. On the GitHub site, users may also report bugs and other
issues, as well as contribute code to the project, which we actively
encourage.

    \section{A Motivating Example: Linear
Regression}\label{a-motivating-example-linear-regression}

To introduce model definition, fitting and posterior analysis, we first
consider a simple Bayesian linear regression model with normal priors
for the parameters. We are interested in predicting outcomes $Y$ as
normally-distributed observations with an expected value $\mu$ that is a
linear function of two predictor variables, $X_1$ and $X_2$.

\[\begin{aligned}
Y  &\sim \mathcal{N}(\mu, \sigma^2) \\
\mu &= \alpha + \beta_1 X_1 + \beta_2 X_2
\end{aligned}\]

where $\alpha$ is the intercept, and $\beta_i$ is the coefficient for
covariate $X_i$, while $\sigma$ represents the observation error. Since
we are constructing a Bayesian model, the unknown variables in the model
must be assigned a prior distribution. We choose zero-mean normal priors
with variance of 100 for both regression coefficients, which corresponds
to \emph{weak} information regarding the true parameter values. We
choose a half-normal distribution (normal distribution bounded at zero)
as the prior for $\sigma$.

\[\begin{aligned}
\alpha &\sim \mathcal{N}(0, 100) \\
\beta_i &\sim \mathcal{N}(0, 100) \\
\sigma &\sim \lvert\mathcal{N}(0, 1){\rvert}
\end{aligned}\]

\subsection{Generating data}\label{generating-data}

We can simulate some artificial data from this model using only NumPy's
\texttt{random} module, and then use PyMC3 to try to recover the
corresponding parameters. We are intentionally generating the data to
closely correspond the PyMC3 model structure.

    \begin{Verbatim}[commandchars=\\\{\}]
{\color{incolor}In [{\color{incolor}1}]:} \PY{k+kn}{import} \PY{n+nn}{numpy} \PY{k+kn}{as} \PY{n+nn}{np}
        \PY{k+kn}{import} \PY{n+nn}{pylab} \PY{k+kn}{as} \PY{n+nn}{pl}

        \PY{c}{\PYZsh{} Intialize random number generator}
        \PY{n}{np}\PY{o}{.}\PY{n}{random}\PY{o}{.}\PY{n}{seed}\PY{p}{(}\PY{l+m+mi}{123}\PY{p}{)}

        \PY{c}{\PYZsh{} True parameter values}
        \PY{n}{alpha}\PY{p}{,} \PY{n}{sigma} \PY{o}{=} \PY{l+m+mi}{1}\PY{p}{,} \PY{l+m+mi}{1}
        \PY{n}{beta} \PY{o}{=} \PY{p}{[}\PY{l+m+mi}{1}\PY{p}{,} \PY{l+m+mf}{2.5}\PY{p}{]}

        \PY{c}{\PYZsh{} Size of dataset}
        \PY{n}{size} \PY{o}{=} \PY{l+m+mi}{100}

        \PY{c}{\PYZsh{} Predictor variable}
        \PY{n}{X1} \PY{o}{=} \PY{n}{np}\PY{o}{.}\PY{n}{linspace}\PY{p}{(}\PY{l+m+mi}{0}\PY{p}{,} \PY{l+m+mi}{1}\PY{p}{,} \PY{n}{size}\PY{p}{)}
        \PY{n}{X2} \PY{o}{=} \PY{n}{np}\PY{o}{.}\PY{n}{linspace}\PY{p}{(}\PY{l+m+mi}{0}\PY{p}{,}\PY{o}{.}\PY{l+m+mi}{2}\PY{p}{,} \PY{n}{size}\PY{p}{)}

        \PY{c}{\PYZsh{} Simulate outcome variable}
        \PY{n}{Y} \PY{o}{=} \PY{n}{alpha} \PY{o}{+} \PY{n}{beta}\PY{p}{[}\PY{l+m+mi}{0}\PY{p}{]}\PY{o}{*}\PY{n}{X1} \PY{o}{+} \PY{n}{beta}\PY{p}{[}\PY{l+m+mi}{1}\PY{p}{]}\PY{o}{*}\PY{n}{X2} \PY{o}{+} \PY{n}{np}\PY{o}{.}\PY{n}{random}\PY{o}{.}\PY{n}{randn}\PY{p}{(}\PY{n}{size}\PY{p}{)}\PY{o}{*}\PY{n}{sigma}
\end{Verbatim}

    Here is what the simulated data look like. We use the \texttt{pylab}
module from the plotting library matplotlib.

    \begin{Verbatim}[commandchars=\\\{\}]
{\color{incolor}In [{\color{incolor}2}]:} \PY{o}{\PYZpc{}}\PY{k}{pylab} inline

        \PY{n}{fig}\PY{p}{,} \PY{n}{axes} \PY{o}{=} \PY{n}{subplots}\PY{p}{(}\PY{l+m+mi}{1}\PY{p}{,} \PY{l+m+mi}{2}\PY{p}{,} \PY{n}{sharex}\PY{o}{=}\PY{n+nb+bp}{True}\PY{p}{,} \PY{n}{figsize}\PY{o}{=}\PY{p}{(}\PY{l+m+mi}{10}\PY{p}{,}\PY{l+m+mi}{4}\PY{p}{)}\PY{p}{)}
        \PY{n}{axes}\PY{p}{[}\PY{l+m+mi}{0}\PY{p}{]}\PY{o}{.}\PY{n}{scatter}\PY{p}{(}\PY{n}{X1}\PY{p}{,} \PY{n}{Y}\PY{p}{)}
        \PY{n}{axes}\PY{p}{[}\PY{l+m+mi}{1}\PY{p}{]}\PY{o}{.}\PY{n}{scatter}\PY{p}{(}\PY{n}{X2}\PY{p}{,} \PY{n}{Y}\PY{p}{)}
        \PY{n}{axes}\PY{p}{[}\PY{l+m+mi}{0}\PY{p}{]}\PY{o}{.}\PY{n}{set\PYZus{}ylabel}\PY{p}{(}\PY{l+s}{\PYZsq{}}\PY{l+s}{Y}\PY{l+s}{\PYZsq{}}\PY{p}{)}\PY{p}{;} \PY{n}{axes}\PY{p}{[}\PY{l+m+mi}{0}\PY{p}{]}\PY{o}{.}\PY{n}{set\PYZus{}xlabel}\PY{p}{(}\PY{l+s}{\PYZsq{}}\PY{l+s}{X1}\PY{l+s}{\PYZsq{}}\PY{p}{)}\PY{p}{;} \PY{n}{axes}\PY{p}{[}\PY{l+m+mi}{1}\PY{p}{]}\PY{o}{.}\PY{n}{set\PYZus{}xlabel}\PY{p}{(}\PY{l+s}{\PYZsq{}}\PY{l+s}{X2}\PY{l+s}{\PYZsq{}}\PY{p}{)}\PY{p}{;}
\end{Verbatim}

    \begin{Verbatim}[commandchars=\\\{\}]
Populating the interactive namespace from numpy and matplotlib
    \end{Verbatim}

    \begin{Verbatim}[commandchars=\\\{\}]
WARNING: pylab import has clobbered these variables: ['beta', 'size']
`\%matplotlib` prevents importing * from pylab and numpy
    \end{Verbatim}

    \begin{center}
    \adjustimage{max size={0.9\linewidth}{0.9\paperheight}}{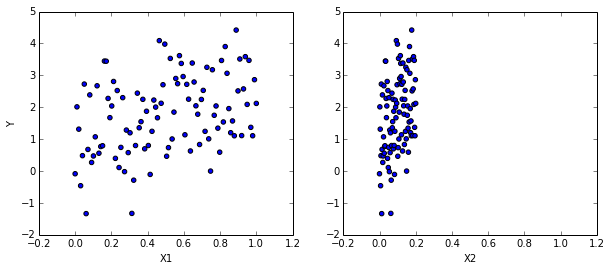}
    \end{center}
    { \hspace*{\fill} \\}

    \subsection{Model Specification}\label{model-specification}

Specifiying this model in PyMC3 is straightforward because the syntax is
as close to the statistical notation. For the most part, each line of
Python code corresponds to a line in the model notation above.

First, we import the components we will need from PyMC.

    \begin{Verbatim}[commandchars=\\\{\}]
{\color{incolor}In [{\color{incolor}3}]:} \PY{k+kn}{from} \PY{n+nn}{pymc3} \PY{k+kn}{import} \PY{n}{Model}\PY{p}{,} \PY{n}{Normal}\PY{p}{,} \PY{n}{HalfNormal}
\end{Verbatim}

    Now we build our model, which we will present in full first, then
explain each part line-by-line.

    \begin{Verbatim}[commandchars=\\\{\}]
{\color{incolor}In [{\color{incolor}4}]:} \PY{n}{basic\PYZus{}model} \PY{o}{=} \PY{n}{Model}\PY{p}{(}\PY{p}{)}

        \PY{k}{with} \PY{n}{basic\PYZus{}model}\PY{p}{:}

            \PY{c}{\PYZsh{} Priors for unknown model parameters}
            \PY{n}{alpha} \PY{o}{=} \PY{n}{Normal}\PY{p}{(}\PY{l+s}{\PYZsq{}}\PY{l+s}{alpha}\PY{l+s}{\PYZsq{}}\PY{p}{,} \PY{n}{mu}\PY{o}{=}\PY{l+m+mi}{0}\PY{p}{,} \PY{n}{sd}\PY{o}{=}\PY{l+m+mi}{10}\PY{p}{)}
            \PY{n}{beta} \PY{o}{=} \PY{n}{Normal}\PY{p}{(}\PY{l+s}{\PYZsq{}}\PY{l+s}{beta}\PY{l+s}{\PYZsq{}}\PY{p}{,} \PY{n}{mu}\PY{o}{=}\PY{l+m+mi}{0}\PY{p}{,} \PY{n}{sd}\PY{o}{=}\PY{l+m+mi}{10}\PY{p}{,} \PY{n}{shape}\PY{o}{=}\PY{l+m+mi}{2}\PY{p}{)}
            \PY{n}{sigma} \PY{o}{=} \PY{n}{HalfNormal}\PY{p}{(}\PY{l+s}{\PYZsq{}}\PY{l+s}{sigma}\PY{l+s}{\PYZsq{}}\PY{p}{,} \PY{n}{sd}\PY{o}{=}\PY{l+m+mi}{1}\PY{p}{)}

            \PY{c}{\PYZsh{} Expected value of outcome}
            \PY{n}{mu} \PY{o}{=} \PY{n}{alpha} \PY{o}{+} \PY{n}{beta}\PY{p}{[}\PY{l+m+mi}{0}\PY{p}{]}\PY{o}{*}\PY{n}{X1} \PY{o}{+} \PY{n}{beta}\PY{p}{[}\PY{l+m+mi}{1}\PY{p}{]}\PY{o}{*}\PY{n}{X2}

            \PY{c}{\PYZsh{} Likelihood (sampling distribution) of observations}
            \PY{n}{Y\PYZus{}obs} \PY{o}{=} \PY{n}{Normal}\PY{p}{(}\PY{l+s}{\PYZsq{}}\PY{l+s}{Y\PYZus{}obs}\PY{l+s}{\PYZsq{}}\PY{p}{,} \PY{n}{mu}\PY{o}{=}\PY{n}{mu}\PY{p}{,} \PY{n}{sd}\PY{o}{=}\PY{n}{sigma}\PY{p}{,} \PY{n}{observed}\PY{o}{=}\PY{n}{Y}\PY{p}{)}
\end{Verbatim}

    The first line,

\begin{Shaded}
\begin{Highlighting}[]
\NormalTok{basic_model = Model()}
\end{Highlighting}
\end{Shaded}

creates a new \texttt{Model} object which is a container for the model
random variables.

Following instantiation of the model, the subsequent specification of
the model components is performed inside a \texttt{with} statement:

\begin{Shaded}
\begin{Highlighting}[]
\KeywordTok{with} \NormalTok{basic_model:}
\end{Highlighting}
\end{Shaded}

This creates a \emph{context manager}, with our \texttt{basic\_model} as
the context, that includes all statements until the indented block ends.
This means all PyMC3 objects introduced in the indented code block below
the \texttt{with} statement are added to the model behind the scenes.
Absent this context manager idiom, we would be forced to manually
associate each of the variables with \texttt{basic\_model} right after
we create them. If you try to create a new random variable without a
\texttt{with model:} statement, it will raise an error since there is no
obvious model for the variable to be added to.

The first three statements in the context manager:

\begin{Shaded}
\begin{Highlighting}[]
\NormalTok{alpha = Normal(}\StringTok{'alpha'}\NormalTok{, mu=}\DecValTok{0}\NormalTok{, sd=}\DecValTok{10}\NormalTok{)}
\NormalTok{beta = Normal(}\StringTok{'beta'}\NormalTok{, mu=}\DecValTok{0}\NormalTok{, sd=}\DecValTok{10}\NormalTok{, shape=}\DecValTok{2}\NormalTok{)}
\NormalTok{sigma = HalfNormal(}\StringTok{'sigma'}\NormalTok{, sd=}\DecValTok{1}\NormalTok{)}
\end{Highlighting}
\end{Shaded}

create a \textbf{stochastic} random variables with a Normal prior
distributions for the regression coefficients with a mean of 0 and
standard deviation of 10 for the regression coefficients, and a
half-normal distribution for the standard deviation of the observations,
$\sigma$. These are stochastic because their values are partly
determined by its parents in the dependency graph of random variables,
which for priors are simple constants, and partly random (or
stochastic).

We call the \texttt{Normal} constructor to create a random variable to
use as a normal prior. The first argument is always the \emph{name} of
the random variable, which should almost always match the name of the
Python variable being assigned to, since it sometimes used to retrieve
the variable from the model for summarizing output. The remaining
required arguments for a stochastic object are the parameters, in this
case \texttt{mu}, the mean, and \texttt{sd}, the standard deviation,
which we assign hyperparameter values for the model. In general, a
distribution's parameters are values that determine the location, shape
or scale of the random variable, depending on the parameterization of
the distribution. Most commonly used distributions, such as
\texttt{Beta}, \texttt{Exponential}, \texttt{Categorical},
\texttt{Gamma}, \texttt{Binomial} and many others, are available in
PyMC3.

The \texttt{beta} variable has an additional \texttt{shape} argument to
denote it as a vector-valued parameter of size 2. The \texttt{shape}
argument is available for all distributions and specifies the length or
shape of the random variable, but is optional for scalar variables,
since it defaults to a value of one. It can be an integer, to specify an
array, or a tuple, to specify a multidimensional array (\emph{e.g.}
\texttt{shape=(5,7)} makes random variable that takes on 5 by 7 matrix
values).

Detailed notes about distributions, sampling methods and other PyMC3
functions are available via the \texttt{help} function.

    \begin{Verbatim}[commandchars=\\\{\}]
{\color{incolor}In [{\color{incolor}5}]:} \PY{n}{help}\PY{p}{(}\PY{n}{Normal}\PY{p}{)} \PY{c}{\PYZsh{}try help(Model), help(Uniform) or help(basic\PYZus{}model)}
\end{Verbatim}

    \begin{Verbatim}[commandchars=\\\{\}]
Help on class Normal in module pymc3.distributions.continuous:

class Normal(pymc3.distributions.distribution.Continuous)
 |  Normal log-likelihood.
 |
 |  .. math::
ight\textbackslash{}\}
 |
 |  Parameters
 |  ----------
 |  mu : float
 |      Mean of the distribution.
 |  tau : float
 |      Precision of the distribution, which corresponds to
 |      :math:`1/\textbackslash{}sigma\^{}2` (tau > 0).
 |  sd : float
 |      Standard deviation of the distribution. Alternative parameterization.
 |
 |  .. note::
 |  - :math:`E(X) = \textbackslash{}mu`
 |  - :math:`Var(X) = 1/        au`
 |
 |  Method resolution order:
 |      Normal
 |      pymc3.distributions.distribution.Continuous
 |      pymc3.distributions.distribution.Distribution
 |      \_\_builtin\_\_.object
 |
 |  Methods defined here:
 |
 |  \_\_init\_\_(self, mu=0.0, tau=None, sd=None, *args, **kwargs)
 |
 |  logp(self, value)
 |
 |  ----------------------------------------------------------------------
 |  Methods inherited from pymc3.distributions.distribution.Distribution:
 |
 |  \_\_getnewargs\_\_(self)
 |
 |  default(self)
 |
 |  get\_test\_val(self, val, defaults)
 |
 |  getattr\_value(self, val)
 |
 |  ----------------------------------------------------------------------
 |  Class methods inherited from pymc3.distributions.distribution.Distribution:
 |
 |  dist(cls, *args, **kwargs) from \_\_builtin\_\_.type
 |
 |  ----------------------------------------------------------------------
 |  Static methods inherited from pymc3.distributions.distribution.Distribution:
 |
 |  \_\_new\_\_(cls, name, *args, **kwargs)
 |
 |  ----------------------------------------------------------------------
 |  Data descriptors inherited from pymc3.distributions.distribution.Distribution:
 |
 |  \_\_dict\_\_
 |      dictionary for instance variables (if defined)
 |
 |  \_\_weakref\_\_
 |      list of weak references to the object (if defined)
    \end{Verbatim}

    Having defined the priors, the next statement creates the expected value
\texttt{mu} of the outcomes, specifying the linear relationship:

\begin{Shaded}
\begin{Highlighting}[]
\NormalTok{mu = alpha + beta[}\DecValTok{0}\NormalTok{]*X1 + beta[}\DecValTok{1}\NormalTok{]*X2}
\end{Highlighting}
\end{Shaded}

This creates a \textbf{deterministic} random variable, which implies
that its value is \emph{completely} determined by its parents' values.
That is, there is no uncertainty beyond that which is inherent in the
parents' values. Here, \texttt{mu} is just the sum of the intercept
\texttt{alpha} and the two products of the coefficients in \texttt{beta}
and the predictor variables, whatever their values may be.

PyMC3 random variables and data can be arbitrarily added, subtracted,
divided, multiplied together and indexed-into to create new random
variables. This allows for great model expressivity. Many common
mathematical functions like \texttt{sum}, \texttt{sin}, \texttt{exp} and
linear algebra functions like \texttt{dot} (for inner product) and
\texttt{inv} (for inverse) are also provided.

The final line of the model, defines \texttt{Y\_obs}, the sampling
distribution of the outcomes in the dataset.

\begin{Shaded}
\begin{Highlighting}[]
\NormalTok{Y_obs = Normal(}\StringTok{'Y_obs'}\NormalTok{, mu=mu, sd=sigma, observed=Y)}
\end{Highlighting}
\end{Shaded}

This is a special case of a stochastic variable that we call an
\textbf{observed stochastic}, and represents the data likelihood of the
model. It is identical to a standard stochastic, except that its
\texttt{observed} argument, which passes the data to the variable,
indicates that the values for this variable were observed, and should
not be changed by any fitting algorithm applied to the model. The data
can be passed in the form of either a \texttt{numpy.ndarray} or
\texttt{pandas.DataFrame} object.

Notice that, unlike for the priors of the model, the parameters for the
normal distribution of \texttt{Y\_obs} are not fixed values, but rather
are the deterministic object \texttt{mu} and the stochastic
\texttt{sigma}. This creates parent-child relationships between the
likelihood and these two variables.

    \subsection{Model fitting}\label{model-fitting}

Having completely specified our model, the next step is to obtain
posterior estimates for the unknown variables in the model. Ideally, we
could calculate the posterior estimates analytically, but for most
non-trivial models, this is not feasible. We will consider two
approaches, whose appropriateness depends on the structure of the model
and the goals of the analysis: finding the \emph{maximum a posteriori}
(MAP) point using optimization methods, and computing summaries based on
samples drawn from the posterior distribution using Markov Chain Monte
Carlo (MCMC) sampling methods.

\paragraph{Maximum a posteriori
methods}\label{maximum-a-posteriori-methods}

The \textbf{maximum a posteriori (MAP)} estimate for a model, is the
mode of the posterior distribution and is generally found using
numerical optimization methods. This is often fast and easy to do, but
only gives a point estimate for the parameters and can be biased if the
mode isn't representative of the distribution. PyMC3 provides this
functionality with the \texttt{find\_MAP} function.

Below we find the MAP for our original model. The MAP is returned as a
parameter \textbf{point}, which is always represented by a Python
dictionary of variable names to NumPy arrays of parameter values.

    \begin{Verbatim}[commandchars=\\\{\}]
{\color{incolor}In [{\color{incolor}6}]:} \PY{k+kn}{from} \PY{n+nn}{pymc3} \PY{k+kn}{import} \PY{n}{find\PYZus{}MAP}

        \PY{n}{map\PYZus{}estimate} \PY{o}{=} \PY{n}{find\PYZus{}MAP}\PY{p}{(}\PY{n}{model}\PY{o}{=}\PY{n}{basic\PYZus{}model}\PY{p}{)}

        \PY{k}{print}\PY{p}{(}\PY{n}{map\PYZus{}estimate}\PY{p}{)}
\end{Verbatim}

    \begin{Verbatim}[commandchars=\\\{\}]
\{'alpha': array(1.0136640995128503), 'beta': array([ 1.46791595,  0.29358319]), 'sigma\_log': array(0.11928764983495602)\}
    \end{Verbatim}

    By default, \texttt{find\_MAP} uses the
Broyden--Fletcher--Goldfarb--Shanno (BFGS) optimization algorithm to
find the maximum of the log-posterior but also allows selection of other
optimization algorithms from the \texttt{scipy.optimize} module. For
example, below we use Powell's method to find the MAP.

    \begin{Verbatim}[commandchars=\\\{\}]
{\color{incolor}In [{\color{incolor}7}]:} \PY{k+kn}{from} \PY{n+nn}{scipy} \PY{k+kn}{import} \PY{n}{optimize}

        \PY{n}{map\PYZus{}estimate} \PY{o}{=} \PY{n}{find\PYZus{}MAP}\PY{p}{(}\PY{n}{model}\PY{o}{=}\PY{n}{basic\PYZus{}model}\PY{p}{,} \PY{n}{fmin}\PY{o}{=}\PY{n}{optimize}\PY{o}{.}\PY{n}{fmin\PYZus{}powell}\PY{p}{)}

        \PY{k}{print}\PY{p}{(}\PY{n}{map\PYZus{}estimate}\PY{p}{)}
\end{Verbatim}

    \begin{Verbatim}[commandchars=\\\{\}]
\{'alpha': array(1.0175522115056725), 'beta': array([ 1.51426781,  0.03520891]), 'sigma\_log': array(0.1181510683418693)\}
    \end{Verbatim}

    It is important to note that the MAP estimate is not always reasonable,
especially if the mode is at an extreme. This can be a subtle issue;
with high dimensional posteriors, one can have areas of extremely high
density but low total probability because the volume is very small. This
will often occur in hierarchical models with the variance parameter for
the random effect. If the individual group means are all the same, the
posterior will have near infinite density if the scale parameter for the
group means is almost zero, even though the probability of such a small
scale parameter will be small since the group means must be extremely
close together.

Most techniques for finding the MAP estimate also only find a
\emph{local} optimium (which is often good enough), but can fail badly
for multimodal posteriors if the different modes are meaningfully
different.

    \paragraph{Sampling methods}\label{sampling-methods}

Though finding the MAP is a fast and easy way of obtaining estimates of
the unknown model parameters, it is limited because there is no
associated estimate of uncertainty produced with the MAP estimates.
Instead, a simulation-based approach such as Markov chain Monte Carlo
(MCMC) can be used to obtain a Markov chain of values that, given the
satisfaction of certain conditions, are indistinguishable from samples
from the posterior distribution.

To conduct MCMC sampling to generate posterior samples in PyMC3, we
specify a \textbf{step method} object that corresponds to a particular
MCMC algorithm, such as Metropolis, Slice sampling, or the No-U-Turn
Sampler (NUTS). PyMC3's \texttt{step\_methods} submodule contains the
following samplers: \texttt{NUTS}, \texttt{Metropolis}, \texttt{Slice},
\texttt{HamiltonianMC}, and \texttt{BinaryMetropolis}.

    \paragraph{Gradient-based sampling
methods}\label{gradient-based-sampling-methods}

PyMC3 has the standard sampling algorithms like adaptive
Metropolis-Hastings and adaptive slice sampling, but PyMC3's most
capable step method is the No-U-Turn Sampler. NUTS is especially useful
on models that have many continuous parameters, a situatiuon where other
MCMC algorithms work very slowly. It takes advantage of information
about where regions of higher probability are, based on the gradient of
the log posterior-density. This helps it achieve dramatically faster
convergence on large problems than traditional sampling methods achieve.
PyMC3 relies on Theano to analytically compute model gradients via
automatic differentation of the posterior density. NUTS also has several
self-tuning strategies for adaptively setting the tunable parameters of
Hamiltonian Monte Carlo. For random variables that are undifferentiable
(namely, discrete variables) NUTS cannot be used, but it may still be
used on the differentiable variables in a model that contains
undifferentiable variables.

NUTS requires a scaling matrix parameter, which is analogous to the
variance parameter for the jump proposal distribution in
Metropolis-Hastings, althrough NUTS uses it somewhat differently. The
matrix gives the rough shape of the distribution so that NUTS does not
make jumps that are too large in some directions and too small in other
directions. It is important to set this scaling parameter to a
reasonable value to facilitate efficient sampling. This is especially
true for models that have many unobserved stochastic random variables or
models with highly non-normal posterior distributions. Poor scaling
parameters will slow down NUTS significantly, sometimes almost stopping
it completely. A reasonable starting point for sampling can also be
important for efficient sampling, but not as often.

Fortunately NUTS can often make good guesses for the scaling parameters.
If you pass a point in parameter space (as a dictionary of variable
names to parameter values, the same format as returned by
\texttt{find\_MAP}) to NUTS, it will look at the local curvature of the
log posterior-density (the diagonal of the Hessian matrix) at that point
to make a guess for a good scaling vector, which often results in a good
value. The MAP estimate is often a good point to use to initiate
sampling. It is also possible to supply your own vector or scaling
matrix to NUTS, though this is a more advanced use. If you wish to
modify a Hessian at a specific point to use as your scaling matrix or
vector, you can use \texttt{find\_hessian} or
\texttt{find\_hessian\_diag}.

For our basic linear regression example in \texttt{basic\_model}, we
will use NUTS to sample 2000 draws from the posterior using the MAP as
the starting point and scaling point. This must also be performed inside
the context of the model.

    \begin{Verbatim}[commandchars=\\\{\}]
{\color{incolor}In [{\color{incolor}8}]:} \PY{k+kn}{from} \PY{n+nn}{pymc3} \PY{k+kn}{import} \PY{n}{NUTS}\PY{p}{,} \PY{n}{sample}

        \PY{k}{with} \PY{n}{basic\PYZus{}model}\PY{p}{:}

            \PY{c}{\PYZsh{} obtain starting values via MAP}
            \PY{n}{start} \PY{o}{=} \PY{n}{find\PYZus{}MAP}\PY{p}{(}\PY{n}{fmin}\PY{o}{=}\PY{n}{optimize}\PY{o}{.}\PY{n}{fmin\PYZus{}powell}\PY{p}{)}

            \PY{c}{\PYZsh{} instantiate sampler}
            \PY{n}{step} \PY{o}{=} \PY{n}{NUTS}\PY{p}{(}\PY{n}{scaling}\PY{o}{=}\PY{n}{start}\PY{p}{)}

            \PY{c}{\PYZsh{} draw 2000 posterior samples}
            \PY{n}{trace} \PY{o}{=} \PY{n}{sample}\PY{p}{(}\PY{l+m+mi}{2000}\PY{p}{,} \PY{n}{step}\PY{p}{,} \PY{n}{start}\PY{o}{=}\PY{n}{start}\PY{p}{)}
\end{Verbatim}

    \begin{Verbatim}[commandchars=\\\{\}]
[-----------------100\%-----------------] 2000 of 2000 complete in 10.4 sec
    \end{Verbatim}

    \begin{Verbatim}[commandchars=\\\{\}]
/Library/Python/2.7/site-packages/theano/scan\_module/scan\_perform\_ext.py:133: RuntimeWarning: numpy.ndarray size changed, may indicate binary incompatibility
  from scan\_perform.scan\_perform import *
    \end{Verbatim}

    The \texttt{sample} function runs the step method(s) passed to it for
the given number of iterations and returns a \texttt{Trace} object
containing the samples collected, in the order they were collected. The
\texttt{trace} object can be queried in a similar way to a \texttt{dict}
containing a map from variable names to \texttt{numpy.array}s. The first
dimension of the array is the sampling index and the later dimensions
match the shape of the variable. We can see the last 5 values for the
\texttt{alpha} variable as follows

    \begin{Verbatim}[commandchars=\\\{\}]
{\color{incolor}In [{\color{incolor}9}]:} \PY{n}{trace}\PY{p}{[}\PY{l+s}{\PYZsq{}}\PY{l+s}{alpha}\PY{l+s}{\PYZsq{}}\PY{p}{]}\PY{p}{[}\PY{o}{\PYZhy{}}\PY{l+m+mi}{5}\PY{p}{:}\PY{p}{]}
\end{Verbatim}

            \begin{Verbatim}[commandchars=\\\{\}]
{\color{outcolor}Out[{\color{outcolor}9}]:} array([ 0.93582185,  0.94703037,  1.0502649 ,  0.91058163,  0.96962219])
\end{Verbatim}

    \subsection{Posterior analysis}\label{posterior-analysis}

\texttt{PyMC3} provides plotting and summarization functions for
inspecting the sampling output. A simple posterior plot can be created
using \texttt{traceplot}.

    \begin{Verbatim}[commandchars=\\\{\}]
{\color{incolor}In [{\color{incolor}10}]:} \PY{k+kn}{from} \PY{n+nn}{pymc3} \PY{k+kn}{import} \PY{n}{traceplot}

         \PY{n}{traceplot}\PY{p}{(}\PY{n}{trace}\PY{p}{)}\PY{p}{;}
\end{Verbatim}

    \begin{center}
    \adjustimage{max size={0.9\linewidth}{0.9\paperheight}}{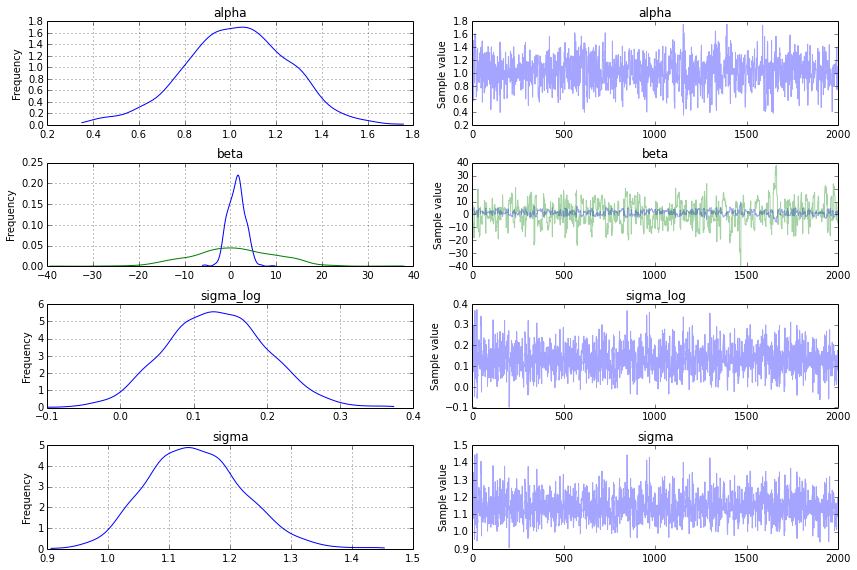}
    \end{center}
    { \hspace*{\fill} \\}

    The left column consists of a smoothed histogram (using kernel density
estimation) of the marginal posteriors of each stochastic random
variable while the right column contains the samples of the Markov chain
plotted in sequential order. The \texttt{beta} variable, being
vector-valued, produces two histograms and two sample traces,
corresponding to both predictor coefficients.

In addition, the \texttt{summary} function provides a text-based output
of common posterior statistics:

    \begin{Verbatim}[commandchars=\\\{\}]
{\color{incolor}In [{\color{incolor}11}]:} \PY{k+kn}{from} \PY{n+nn}{pymc3} \PY{k+kn}{import} \PY{n}{summary}

         \PY{n}{summary}\PY{p}{(}\PY{n}{trace}\PY{p}{)}
\end{Verbatim}

    \begin{Verbatim}[commandchars=\\\{\}]
alpha:

  Mean             SD               MC Error         95\% HPD interval
  -------------------------------------------------------------------

  1.027            0.227            0.007            [0.592, 1.487]

  Posterior quantiles:
  2.5            25             50             75             97.5
  |--------------|==============|==============|--------------|

  0.552          0.878          1.031          1.185          1.456


beta:

  Mean             SD               MC Error         95\% HPD interval
  -------------------------------------------------------------------

  1.340            1.861            0.099            [-1.997, 4.799]
  0.838            9.188            0.488            [-16.926, 16.472]

  Posterior quantiles:
  2.5            25             50             75             97.5
  |--------------|==============|==============|--------------|

  -1.994         0.027          1.413          2.540          4.805
  -16.534        -5.052         0.713          7.316          16.939


sigma\_log:

  Mean             SD               MC Error         95\% HPD interval
  -------------------------------------------------------------------

  0.132            0.068            0.002            [-0.002, 0.258]

  Posterior quantiles:
  2.5            25             50             75             97.5
  |--------------|==============|==============|--------------|

  0.002          0.084          0.130          0.176          0.269


sigma:

  Mean             SD               MC Error         95\% HPD interval
  -------------------------------------------------------------------

  1.143            0.079            0.002            [0.998, 1.294]

  Posterior quantiles:
  2.5            25             50             75             97.5
  |--------------|==============|==============|--------------|

  1.002          1.088          1.139          1.193          1.308
    \end{Verbatim}

    \section{Case study 1: Stochastic
volatility}\label{case-study-1-stochastic-volatility}

We present a case study of stochastic volatility, time varying stock
market volatility, to illustrate PyMC3's use in addressing a more
realistic problem. The distribution of market returns is highly
non-normal, which makes sampling the volatlities significantly more
difficult. This example has 400+ parameters so using common sampling
algorithms like Metropolis-Hastings would get bogged down, generating
highly autocorrelated samples. Instead, we use NUTS, which is
dramatically more efficient.

\subsection{The Model}\label{the-model}

Asset prices have time-varying volatility (variance of day over day
\texttt{returns}). In some periods, returns are highly variable, while
in others they are very stable. Stochastic volatility models address
this with a latent volatility variable, which changes over time. The
following model is similar to the one described in the NUTS paper
(Hoffman 2014, p.~21).

\[\begin{aligned}
  \sigma &\sim exp(50) \\
  \nu &\sim exp(.1) \\
  s_i &\sim \mathcal{N}(s_{i-1}, \sigma^{-2}) \\
  log(y_i) &\sim t(\nu, 0, exp(-2 s_i))
\end{aligned}\]

Here, $y$ is the daily return series which is modeled with a Student-t
distribution with an unknown degrees of freedom parameter, and a scale
parameter determined by a latent process $s$. The individual $s_i$ are
the individual daily log volatilities in the latent log volatility
process.

    \subsection{The Data}\label{the-data}

Our data consist of the last 400 daily returns of the S\&P 500.

    \begin{Verbatim}[commandchars=\\\{\}]
{\color{incolor}In [{\color{incolor}12}]:} \PY{n}{n} \PY{o}{=} \PY{l+m+mi}{400}
         \PY{n}{returns} \PY{o}{=} \PY{n}{np}\PY{o}{.}\PY{n}{genfromtxt}\PY{p}{(}\PY{l+s}{\PYZdq{}}\PY{l+s}{data/SP500.csv}\PY{l+s}{\PYZdq{}}\PY{p}{)}\PY{p}{[}\PY{o}{\PYZhy{}}\PY{n}{n}\PY{p}{:}\PY{p}{]}
         \PY{n}{pl}\PY{o}{.}\PY{n}{plot}\PY{p}{(}\PY{n}{returns}\PY{p}{)}\PY{p}{;}
\end{Verbatim}

    \begin{center}
    \adjustimage{max size={0.9\linewidth}{0.9\paperheight}}{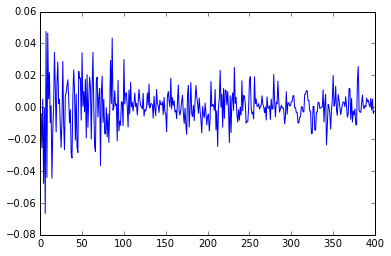}
    \end{center}
    { \hspace*{\fill} \\}

    \subsection{Model Specification}\label{model-specification}

As with the linear regession example, specifying the model in PyMC3
mirrors its statistical specification. This model employs several new
distributions: the \texttt{Exponential} distribution for the $\nu$
and $\sigma$ priors, the student-t (\texttt{T}) distribution for
distribution of returns, and the \texttt{GaussianRandomWalk} for the
prior for the latent volatilities.

In PyMC3, variables with purely positive priors like
\texttt{Exponential} are transformed with a log transform. This makes
sampling more robust. Behind the scenes, a variable in the unconstrained
space (named ``variableName\_log'') is added to the model for sampling.
In this model his happens behind the scenes for both the degrees of
freedom, \texttt{nu}, and the scale parameter for the volatility
process, \texttt{sigma}, since they both have exponential priors.
Variables with priors that constrain them on two sides, like
\texttt{Beta} or \texttt{Uniform}, are also transformed to be
unconstrained but with a log odds transform.

Although, unlike model specifiation in PyMC2, we do not typically
provide starting points for variables at the model specification stage,
we can also provide an initial value for any distribution (called a
``test value'') using the \texttt{testval} argument. This overrides the
default test value for the distribution (usually the mean, median or
mode of the distribution), and is most often useful if some values are
illegal and we want to ensure we select a legal one. The test values for
the distributions are also used as a starting point for sampling and
optimization by default, though this is easily overriden.

The vector of latent volatilities \texttt{s} is given a prior
distribution by \texttt{GaussianRandomWalk}. As its name suggests
GaussianRandomWalk is a vector valued distribution where the values of
the vector form a random normal walk of length n, as specified by the
\texttt{shape} argument. The scale of the innovations of the random
walk, \texttt{sigma}, is specified in terms of the precision of the
normally distributed innovations and can be a scalar or vector.

    \begin{Verbatim}[commandchars=\\\{\}]
{\color{incolor}In [{\color{incolor}14}]:} \PY{k+kn}{from} \PY{n+nn}{pymc3} \PY{k+kn}{import} \PY{n}{Exponential}\PY{p}{,} \PY{n}{T}\PY{p}{,} \PY{n}{exp}\PY{p}{,} \PY{n}{Deterministic}
         \PY{k+kn}{from} \PY{n+nn}{pymc3.distributions.timeseries} \PY{k+kn}{import} \PY{n}{GaussianRandomWalk}

         \PY{k}{with} \PY{n}{Model}\PY{p}{(}\PY{p}{)} \PY{k}{as} \PY{n}{sp500\PYZus{}model}\PY{p}{:}

             \PY{n}{nu} \PY{o}{=} \PY{n}{Exponential}\PY{p}{(}\PY{l+s}{\PYZsq{}}\PY{l+s}{nu}\PY{l+s}{\PYZsq{}}\PY{p}{,} \PY{l+m+mf}{1.}\PY{o}{/}\PY{l+m+mi}{10}\PY{p}{,} \PY{n}{testval}\PY{o}{=}\PY{o}{.}\PY{l+m+mi}{1}\PY{p}{)}

             \PY{n}{sigma} \PY{o}{=} \PY{n}{Exponential}\PY{p}{(}\PY{l+s}{\PYZsq{}}\PY{l+s}{sigma}\PY{l+s}{\PYZsq{}}\PY{p}{,} \PY{l+m+mf}{1.}\PY{o}{/}\PY{o}{.}\PY{l+m+mo}{02}\PY{p}{,} \PY{n}{testval}\PY{o}{=}\PY{o}{.}\PY{l+m+mi}{1}\PY{p}{)}

             \PY{n}{s} \PY{o}{=} \PY{n}{GaussianRandomWalk}\PY{p}{(}\PY{l+s}{\PYZsq{}}\PY{l+s}{s}\PY{l+s}{\PYZsq{}}\PY{p}{,} \PY{n}{sigma}\PY{o}{*}\PY{o}{*}\PY{o}{\PYZhy{}}\PY{l+m+mi}{2}\PY{p}{,} \PY{n}{shape}\PY{o}{=}\PY{n}{n}\PY{p}{)}

             \PY{n}{volatility\PYZus{}process} \PY{o}{=} \PY{n}{Deterministic}\PY{p}{(}\PY{l+s}{\PYZsq{}}\PY{l+s}{volatility\PYZus{}process}\PY{l+s}{\PYZsq{}}\PY{p}{,} \PY{n}{exp}\PY{p}{(}\PY{o}{\PYZhy{}}\PY{l+m+mi}{2}\PY{o}{*}\PY{n}{s}\PY{p}{)}\PY{p}{)}

             \PY{n}{r} \PY{o}{=} \PY{n}{T}\PY{p}{(}\PY{l+s}{\PYZsq{}}\PY{l+s}{r}\PY{l+s}{\PYZsq{}}\PY{p}{,} \PY{n}{nu}\PY{p}{,} \PY{n}{lam}\PY{o}{=}\PY{l+m+mi}{1}\PY{o}{/}\PY{n}{volatility\PYZus{}process}\PY{p}{,} \PY{n}{observed}\PY{o}{=}\PY{n}{returns}\PY{p}{)}
\end{Verbatim}

    Notice that we transform the log volatility process \texttt{s} into the
volatility process by \texttt{exp(-2*s)}. Here, \texttt{exp} is a Theano
function, rather than the corresponding function in NumPy; Theano
provides a large subset of the mathematical functions that NumPy does.

Also note that we have declared the \texttt{Model} name
\texttt{sp500\_model} in the first occurrence of the context manager,
rather than splitting it into two lines, as we did for the first
example.

    \subsection{Fitting}\label{fitting}

Before we draw samples from the posterior, it is prudent to find a
decent starting valuwa by finding a point of relatively high
probability. For this model, the full \emph{maximum a posteriori} (MAP)
point over all variables is degenerate and has infinite density. But, if
we fix \texttt{log\_sigma} and \texttt{nu} it is no longer degenerate,
so we find the MAP with respect only to the volatility process
\texttt{s} keeping \texttt{log\_sigma} and \texttt{nu} constant at their
default values (remember that we set \texttt{testval=.1} for
\texttt{sigma}). We use the Limited-memory BFGS (L-BFGS) optimizer,
which is provided by the \texttt{scipy.optimize} package, as it is more
efficient for high dimensional functions and we have 400 stochastic
random variables (mostly from \texttt{s}).

To do the sampling, we do a short initial run to put us in a volume of
high probability, then start again at the new starting point.
\texttt{trace{[}-1{]}} gives us the last point in the sampling trace.
NUTS will recalculate the scaling parameters based on the new point, and
in this case it leads to faster sampling due to better scaling.

    \begin{Verbatim}[commandchars=\\\{\}]
{\color{incolor}In [{\color{incolor}14}]:} \PY{k+kn}{import} \PY{n+nn}{scipy}
         \PY{k}{with} \PY{n}{sp500\PYZus{}model}\PY{p}{:}
             \PY{n}{start} \PY{o}{=} \PY{n}{find\PYZus{}MAP}\PY{p}{(}\PY{n+nb}{vars}\PY{o}{=}\PY{p}{[}\PY{n}{s}\PY{p}{]}\PY{p}{,} \PY{n}{fmin}\PY{o}{=}\PY{n}{scipy}\PY{o}{.}\PY{n}{optimize}\PY{o}{.}\PY{n}{fmin\PYZus{}l\PYZus{}bfgs\PYZus{}b}\PY{p}{)}

             \PY{n}{step} \PY{o}{=} \PY{n}{NUTS}\PY{p}{(}\PY{n}{scaling}\PY{o}{=}\PY{n}{start}\PY{p}{)}
             \PY{n}{trace} \PY{o}{=} \PY{n}{sample}\PY{p}{(}\PY{l+m+mi}{50}\PY{p}{,} \PY{n}{step}\PY{p}{,} \PY{n}{progressbar}\PY{o}{=}\PY{n+nb+bp}{False}\PY{p}{)}

             \PY{c}{\PYZsh{} Start next run at the last sampled position.}
             \PY{n}{step} \PY{o}{=} \PY{n}{NUTS}\PY{p}{(}\PY{n}{scaling}\PY{o}{=}\PY{n}{trace}\PY{p}{[}\PY{o}{\PYZhy{}}\PY{l+m+mi}{1}\PY{p}{]}\PY{p}{,} \PY{n}{gamma}\PY{o}{=}\PY{o}{.}\PY{l+m+mi}{25}\PY{p}{)}
             \PY{n}{trace} \PY{o}{=} \PY{n}{sample}\PY{p}{(}\PY{l+m+mi}{2000}\PY{p}{,} \PY{n}{step}\PY{p}{,} \PY{n}{start}\PY{o}{=}\PY{n}{trace}\PY{p}{[}\PY{o}{\PYZhy{}}\PY{l+m+mi}{1}\PY{p}{]}\PY{p}{,}\PY{n}{progressbar}\PY{o}{=}\PY{n+nb+bp}{False}\PY{p}{,} \PY{n}{njobs}\PY{o}{=}\PY{l+m+mi}{4}\PY{p}{)}
\end{Verbatim}

    We can check our samples by looking at the traceplot for \texttt{nu} and
\texttt{sigma}.

    \begin{Verbatim}[commandchars=\\\{\}]
{\color{incolor}In [{\color{incolor}16}]:} \PY{c}{\PYZsh{}figsize(12,6)}
         \PY{n}{traceplot}\PY{p}{(}\PY{n}{trace}\PY{p}{,} \PY{p}{[}\PY{n}{nu}\PY{p}{,} \PY{n}{sigma}\PY{p}{]}\PY{p}{)}\PY{p}{;}
\end{Verbatim}

    \begin{center}
    \adjustimage{max size={0.9\linewidth}{0.9\paperheight}}{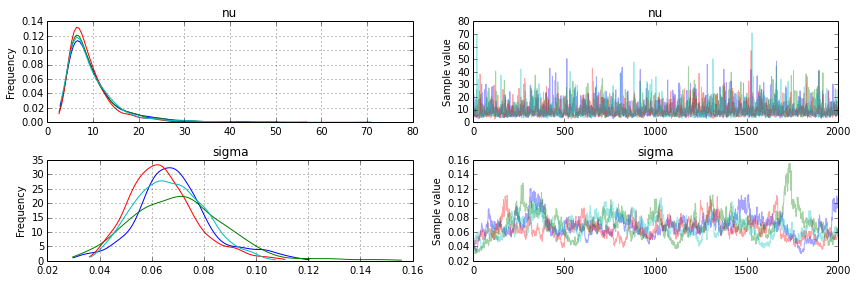}
    \end{center}
    { \hspace*{\fill} \\}

    Finally we plot the distribution of volatility paths by plotting many of
our sampled volatility paths on the same graph. Each is rendered
partially transparent (via the \texttt{alpha} argument in Matplotlib's
\texttt{plot} function) so the regions where many paths overlap are
shaded more darkly.

    \begin{Verbatim}[commandchars=\\\{\}]
{\color{incolor}In [{\color{incolor}16}]:} \PY{n}{figsize}\PY{p}{(}\PY{l+m+mi}{15}\PY{p}{,}\PY{l+m+mi}{8}\PY{p}{)}
         \PY{n}{pl}\PY{o}{.}\PY{n}{title}\PY{p}{(}\PY{l+s}{\PYZdq{}}\PY{l+s}{volatility\PYZus{}process}\PY{l+s}{\PYZdq{}}\PY{p}{)}\PY{p}{;}
         \PY{n}{pl}\PY{o}{.}\PY{n}{plot}\PY{p}{(}\PY{n}{trace}\PY{p}{[}\PY{l+s}{\PYZsq{}}\PY{l+s}{volatility\PYZus{}process}\PY{l+s}{\PYZsq{}}\PY{p}{,}\PY{p}{:}\PY{p}{:}\PY{l+m+mi}{30}\PY{p}{]}\PY{o}{.}\PY{n}{T}\PY{p}{,}\PY{l+s}{\PYZsq{}}\PY{l+s}{b}\PY{l+s}{\PYZsq{}}\PY{p}{,} \PY{n}{alpha}\PY{o}{=}\PY{o}{.}\PY{l+m+mo}{03}\PY{p}{)}\PY{p}{;}
         \PY{n}{pl}\PY{o}{.}\PY{n}{ylim}\PY{p}{(}\PY{l+m+mi}{0}\PY{p}{,} \PY{o}{.}\PY{l+m+mo}{001}\PY{p}{)}
         \PY{n}{pl}\PY{o}{.}\PY{n}{xlabel}\PY{p}{(}\PY{l+s}{\PYZsq{}}\PY{l+s}{time}\PY{l+s}{\PYZsq{}}\PY{p}{)}\PY{p}{;}
         \PY{n}{pl}\PY{o}{.}\PY{n}{ylabel}\PY{p}{(}\PY{l+s}{\PYZsq{}}\PY{l+s}{log volatility}\PY{l+s}{\PYZsq{}}\PY{p}{)}\PY{p}{;}
         \PY{n}{pl}\PY{o}{.}\PY{n}{plot}\PY{p}{(}\PY{n}{returns}\PY{o}{.}\PY{n}{cumsum}\PY{p}{(}\PY{p}{)}\PY{o}{*}\PY{o}{.}\PY{l+m+mo}{0025}\PY{o}{+}\PY{o}{.}\PY{l+m+mo}{0005}\PY{p}{)}\PY{p}{;}
\end{Verbatim}

    \begin{center}
    \adjustimage{max size={0.9\linewidth}{0.9\paperheight}}{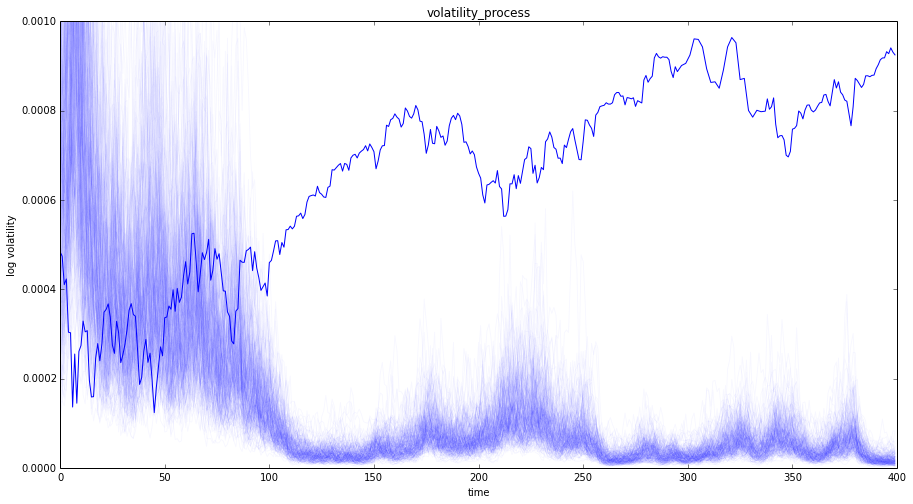}
    \end{center}
    { \hspace*{\fill} \\}

    \section{Case study 2: Coal mining
disasters}\label{case-study-2-coal-mining-disasters}

Consider the following time series of recorded coal mining disasters in
the UK from 1851 to 1962 (Jarrett, 1979). The number of disasters is
thought to have been affected by changes in safety regulations during
this period. Unfortunately, we also have pair of years with missing
data, identified as missing by a NumPy MaskedArray using -999 as the
marker value.

Next we will build a model for this series and attempt to estimate when
the change occured. At the same time, we will see how to handle missing
data, use multiple samplers and sample from discrete random variables.

    \begin{Verbatim}[commandchars=\\\{\}]
{\color{incolor}In [{\color{incolor}17}]:} \PY{n}{disaster\PYZus{}data} \PY{o}{=} \PY{n}{np}\PY{o}{.}\PY{n}{ma}\PY{o}{.}\PY{n}{masked\PYZus{}values}\PY{p}{(}\PY{p}{[}\PY{l+m+mi}{4}\PY{p}{,} \PY{l+m+mi}{5}\PY{p}{,} \PY{l+m+mi}{4}\PY{p}{,} \PY{l+m+mi}{0}\PY{p}{,} \PY{l+m+mi}{1}\PY{p}{,} \PY{l+m+mi}{4}\PY{p}{,} \PY{l+m+mi}{3}\PY{p}{,} \PY{l+m+mi}{4}\PY{p}{,} \PY{l+m+mi}{0}\PY{p}{,} \PY{l+m+mi}{6}\PY{p}{,} \PY{l+m+mi}{3}\PY{p}{,} \PY{l+m+mi}{3}\PY{p}{,} \PY{l+m+mi}{4}\PY{p}{,} \PY{l+m+mi}{0}\PY{p}{,} \PY{l+m+mi}{2}\PY{p}{,} \PY{l+m+mi}{6}\PY{p}{,}
                                     \PY{l+m+mi}{3}\PY{p}{,} \PY{l+m+mi}{3}\PY{p}{,} \PY{l+m+mi}{5}\PY{p}{,} \PY{l+m+mi}{4}\PY{p}{,} \PY{l+m+mi}{5}\PY{p}{,} \PY{l+m+mi}{3}\PY{p}{,} \PY{l+m+mi}{1}\PY{p}{,} \PY{l+m+mi}{4}\PY{p}{,} \PY{l+m+mi}{4}\PY{p}{,} \PY{l+m+mi}{1}\PY{p}{,} \PY{l+m+mi}{5}\PY{p}{,} \PY{l+m+mi}{5}\PY{p}{,} \PY{l+m+mi}{3}\PY{p}{,} \PY{l+m+mi}{4}\PY{p}{,} \PY{l+m+mi}{2}\PY{p}{,} \PY{l+m+mi}{5}\PY{p}{,}
                                     \PY{l+m+mi}{2}\PY{p}{,} \PY{l+m+mi}{2}\PY{p}{,} \PY{l+m+mi}{3}\PY{p}{,} \PY{l+m+mi}{4}\PY{p}{,} \PY{l+m+mi}{2}\PY{p}{,} \PY{l+m+mi}{1}\PY{p}{,} \PY{l+m+mi}{3}\PY{p}{,} \PY{o}{\PYZhy{}}\PY{l+m+mi}{999}\PY{p}{,} \PY{l+m+mi}{2}\PY{p}{,} \PY{l+m+mi}{1}\PY{p}{,} \PY{l+m+mi}{1}\PY{p}{,} \PY{l+m+mi}{1}\PY{p}{,} \PY{l+m+mi}{1}\PY{p}{,} \PY{l+m+mi}{3}\PY{p}{,} \PY{l+m+mi}{0}\PY{p}{,} \PY{l+m+mi}{0}\PY{p}{,}
                                     \PY{l+m+mi}{1}\PY{p}{,} \PY{l+m+mi}{0}\PY{p}{,} \PY{l+m+mi}{1}\PY{p}{,} \PY{l+m+mi}{1}\PY{p}{,} \PY{l+m+mi}{0}\PY{p}{,} \PY{l+m+mi}{0}\PY{p}{,} \PY{l+m+mi}{3}\PY{p}{,} \PY{l+m+mi}{1}\PY{p}{,} \PY{l+m+mi}{0}\PY{p}{,} \PY{l+m+mi}{3}\PY{p}{,} \PY{l+m+mi}{2}\PY{p}{,} \PY{l+m+mi}{2}\PY{p}{,} \PY{l+m+mi}{0}\PY{p}{,} \PY{l+m+mi}{1}\PY{p}{,} \PY{l+m+mi}{1}\PY{p}{,} \PY{l+m+mi}{1}\PY{p}{,}
                                     \PY{l+m+mi}{0}\PY{p}{,} \PY{l+m+mi}{1}\PY{p}{,} \PY{l+m+mi}{0}\PY{p}{,} \PY{l+m+mi}{1}\PY{p}{,} \PY{l+m+mi}{0}\PY{p}{,} \PY{l+m+mi}{0}\PY{p}{,} \PY{l+m+mi}{0}\PY{p}{,} \PY{l+m+mi}{2}\PY{p}{,} \PY{l+m+mi}{1}\PY{p}{,} \PY{l+m+mi}{0}\PY{p}{,} \PY{l+m+mi}{0}\PY{p}{,} \PY{l+m+mi}{0}\PY{p}{,} \PY{l+m+mi}{1}\PY{p}{,} \PY{l+m+mi}{1}\PY{p}{,} \PY{l+m+mi}{0}\PY{p}{,} \PY{l+m+mi}{2}\PY{p}{,}
                                     \PY{l+m+mi}{3}\PY{p}{,} \PY{l+m+mi}{3}\PY{p}{,} \PY{l+m+mi}{1}\PY{p}{,} \PY{o}{\PYZhy{}}\PY{l+m+mi}{999}\PY{p}{,} \PY{l+m+mi}{2}\PY{p}{,} \PY{l+m+mi}{1}\PY{p}{,} \PY{l+m+mi}{1}\PY{p}{,} \PY{l+m+mi}{1}\PY{p}{,} \PY{l+m+mi}{1}\PY{p}{,} \PY{l+m+mi}{2}\PY{p}{,} \PY{l+m+mi}{4}\PY{p}{,} \PY{l+m+mi}{2}\PY{p}{,} \PY{l+m+mi}{0}\PY{p}{,} \PY{l+m+mi}{0}\PY{p}{,} \PY{l+m+mi}{1}\PY{p}{,} \PY{l+m+mi}{4}\PY{p}{,}
                                     \PY{l+m+mi}{0}\PY{p}{,} \PY{l+m+mi}{0}\PY{p}{,} \PY{l+m+mi}{0}\PY{p}{,} \PY{l+m+mi}{1}\PY{p}{,} \PY{l+m+mi}{0}\PY{p}{,} \PY{l+m+mi}{0}\PY{p}{,} \PY{l+m+mi}{0}\PY{p}{,} \PY{l+m+mi}{0}\PY{p}{,} \PY{l+m+mi}{0}\PY{p}{,} \PY{l+m+mi}{1}\PY{p}{,} \PY{l+m+mi}{0}\PY{p}{,} \PY{l+m+mi}{0}\PY{p}{,} \PY{l+m+mi}{1}\PY{p}{,} \PY{l+m+mi}{0}\PY{p}{,} \PY{l+m+mi}{1}\PY{p}{]}\PY{p}{,} \PY{n}{value}\PY{o}{=}\PY{o}{\PYZhy{}}\PY{l+m+mi}{999}\PY{p}{)}
         \PY{n}{year} \PY{o}{=} \PY{n}{np}\PY{o}{.}\PY{n}{arange}\PY{p}{(}\PY{l+m+mi}{1851}\PY{p}{,} \PY{l+m+mi}{1962}\PY{p}{)}

         \PY{n}{plot}\PY{p}{(}\PY{n}{year}\PY{p}{,} \PY{n}{disaster\PYZus{}data}\PY{p}{,} \PY{l+s}{\PYZsq{}}\PY{l+s}{o}\PY{l+s}{\PYZsq{}}\PY{p}{,} \PY{n}{markersize}\PY{o}{=}\PY{l+m+mi}{8}\PY{p}{)}\PY{p}{;}
         \PY{n}{ylabel}\PY{p}{(}\PY{l+s}{\PYZdq{}}\PY{l+s}{Disaster count}\PY{l+s}{\PYZdq{}}\PY{p}{)}
         \PY{n}{xlabel}\PY{p}{(}\PY{l+s}{\PYZdq{}}\PY{l+s}{Year}\PY{l+s}{\PYZdq{}}\PY{p}{)}
\end{Verbatim}

            \begin{Verbatim}[commandchars=\\\{\}]
{\color{outcolor}Out[{\color{outcolor}17}]:} <matplotlib.text.Text at 0x1129c3650>
\end{Verbatim}

    \begin{center}
    \adjustimage{max size={0.9\linewidth}{0.9\paperheight}}{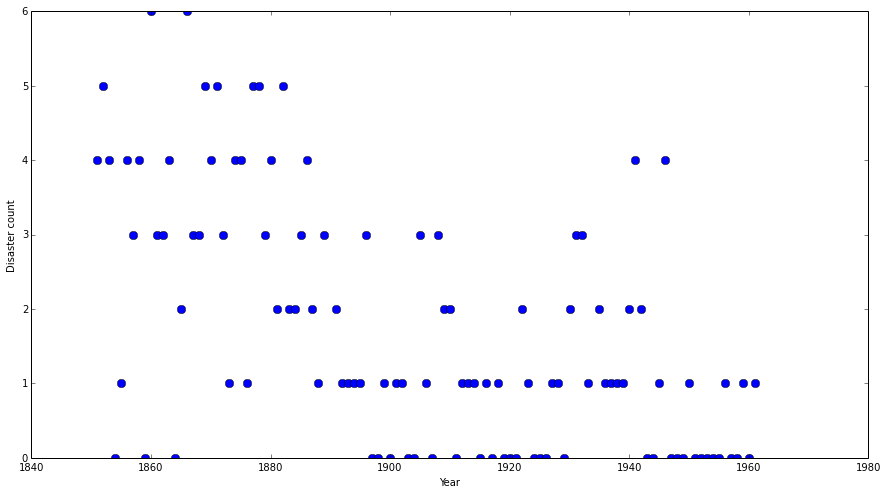}
    \end{center}
    { \hspace*{\fill} \\}

    Occurrences of disasters in the time series is thought to follow a
Poisson process with a large rate parameter in the early part of the
time series, and from one with a smaller rate in the later part. We are
interested in locating the change point in the series, which perhaps is
related to changes in mining safety regulations.

In our model,

\[
\begin{aligned}
  D_t &\sim \text{Pois}(r_t), r_t= \begin{cases}
   l, & \text{if } t \lt s \\
   r, & \text{if } t \ge s
   \end{cases} \\
  s &\sim \text{Unif}(t_l, t_h)\\
  e &\sim \text{exp}(1)\\
  l &\sim \text{exp}(1)
\end{aligned}
\] the parameters are defined as follows: * $D_t$: The number of
disasters in year $t$ * $r_t$: The rate parameter of the Poisson
distribution of disasters in year $t$. * $s$: The year in which the rate
parameter changes (the switchpoint). * $e$: The rate parameter before
the switchpoint $s$. * $l$: The rate parameter after the switchpoint
$s$. * $t_l$, $t_h$: The lower and upper boundaries of year $t$.

This model is built much like our previous models. The major differences
are the introduction of discrete variables with the Poisson and
discrete-uniform priors and the novel form of the deterministic random
variable \texttt{rate}.

    \begin{Verbatim}[commandchars=\\\{\}]
{\color{incolor}In [{\color{incolor}18}]:} \PY{k+kn}{from} \PY{n+nn}{pymc3} \PY{k+kn}{import} \PY{n}{DiscreteUniform}\PY{p}{,} \PY{n}{Poisson}\PY{p}{,} \PY{n}{switch}

         \PY{k}{with} \PY{n}{Model}\PY{p}{(}\PY{p}{)} \PY{k}{as} \PY{n}{disaster\PYZus{}model}\PY{p}{:}

             \PY{n}{switchpoint} \PY{o}{=} \PY{n}{DiscreteUniform}\PY{p}{(}\PY{l+s}{\PYZsq{}}\PY{l+s}{switchpoint}\PY{l+s}{\PYZsq{}}\PY{p}{,} \PY{n}{lower}\PY{o}{=}\PY{n}{year}\PY{o}{.}\PY{n}{min}\PY{p}{(}\PY{p}{)}\PY{p}{,} \PY{n}{upper}\PY{o}{=}\PY{n}{year}\PY{o}{.}\PY{n}{max}\PY{p}{(}\PY{p}{)}\PY{p}{,} \PY{n}{testval}\PY{o}{=}\PY{l+m+mi}{1900}\PY{p}{)}

             \PY{c}{\PYZsh{} Priors for pre\PYZhy{} and post\PYZhy{}switch rates number of disasters}
             \PY{n}{early\PYZus{}rate} \PY{o}{=} \PY{n}{Exponential}\PY{p}{(}\PY{l+s}{\PYZsq{}}\PY{l+s}{early\PYZus{}rate}\PY{l+s}{\PYZsq{}}\PY{p}{,} \PY{l+m+mi}{1}\PY{p}{)}
             \PY{n}{late\PYZus{}rate} \PY{o}{=} \PY{n}{Exponential}\PY{p}{(}\PY{l+s}{\PYZsq{}}\PY{l+s}{late\PYZus{}rate}\PY{l+s}{\PYZsq{}}\PY{p}{,} \PY{l+m+mi}{1}\PY{p}{)}

             \PY{c}{\PYZsh{} Allocate appropriate Poisson rates to years before and after current}
             \PY{n}{rate} \PY{o}{=} \PY{n}{switch}\PY{p}{(}\PY{n}{switchpoint} \PY{o}{\PYZgt{}}\PY{o}{=} \PY{n}{year}\PY{p}{,} \PY{n}{early\PYZus{}rate}\PY{p}{,} \PY{n}{late\PYZus{}rate}\PY{p}{)}

             \PY{n}{disasters} \PY{o}{=} \PY{n}{Poisson}\PY{p}{(}\PY{l+s}{\PYZsq{}}\PY{l+s}{disasters}\PY{l+s}{\PYZsq{}}\PY{p}{,} \PY{n}{rate}\PY{p}{,} \PY{n}{observed}\PY{o}{=}\PY{n}{disaster\PYZus{}data}\PY{p}{)}
\end{Verbatim}

    The logic for the rate random variable,

\begin{Shaded}
\begin{Highlighting}[]
\NormalTok{rate = switch(switchpoint >= year, early_rate, late_rate)}
\end{Highlighting}
\end{Shaded}

is implemented using \texttt{switch}, a Theano function that works like
an if statement. It uses the first argument to switch between the next
two arguments.

Missing values are handled transparently by passing a
\texttt{MaskedArray} or a \texttt{pandas.DataFrame} with NaN values to
the \texttt{observed} argument when creating an observed stochastic
random variable. Behind the scenes, another random variable,
\texttt{disasters.missing\_values} is created to model the missing
values. All we need to do to handle the missing values is ensure we
sample this random variable as well.

    Unfortunately because they are discrete variables and thus have no
meaningful gradient, we cannot use NUTS for sampling
\texttt{switchpoint} or the missing disaster observations. Instead, we
will sample using a \texttt{Metroplis} step method, which implements
adaptive Metropolis-Hastings, because it is designed to handle discrete
values.

We sample with both samplers at once by passing them to the
\texttt{sample} function in a list. Each new sample is generated by
first applying \texttt{step1} then \texttt{step2}.

    \begin{Verbatim}[commandchars=\\\{\}]
{\color{incolor}In [{\color{incolor}19}]:} \PY{k+kn}{from} \PY{n+nn}{pymc3} \PY{k+kn}{import} \PY{n}{Metropolis}

         \PY{k}{with} \PY{n}{disaster\PYZus{}model}\PY{p}{:}
             \PY{n}{step1} \PY{o}{=} \PY{n}{NUTS}\PY{p}{(}\PY{p}{[}\PY{n}{early\PYZus{}rate}\PY{p}{,} \PY{n}{late\PYZus{}rate}\PY{p}{]}\PY{p}{)}

             \PY{c}{\PYZsh{} Use Metropolis for switchpoint, and missing values since it accomodates discrete variables}
             \PY{n}{step2} \PY{o}{=} \PY{n}{Metropolis}\PY{p}{(}\PY{p}{[}\PY{n}{switchpoint}\PY{p}{,} \PY{n}{disasters}\PY{o}{.}\PY{n}{missing\PYZus{}values}\PY{p}{[}\PY{l+m+mi}{0}\PY{p}{]}\PY{p}{]} \PY{p}{)}

             \PY{n}{trace} \PY{o}{=} \PY{n}{sample}\PY{p}{(}\PY{l+m+mi}{10000}\PY{p}{,} \PY{n}{step}\PY{o}{=}\PY{p}{[}\PY{n}{step1}\PY{p}{,} \PY{n}{step2}\PY{p}{]}\PY{p}{)}
\end{Verbatim}

    \begin{Verbatim}[commandchars=\\\{\}]
[-----------------100\%-----------------] 10000 of 10000 complete in 10.3 sec
    \end{Verbatim}

    In the trace plot below we can see that there's about a 10 year span
that's plausible for a significant change in safety, but a 5 year span
that contains most of the probability mass. The distribution is jagged
because of the jumpy relationship beween the year switchpoint and the
likelihood and not due to sampling error.

    \begin{Verbatim}[commandchars=\\\{\}]
{\color{incolor}In [{\color{incolor}21}]:} \PY{n}{traceplot}\PY{p}{(}\PY{n}{trace}\PY{p}{)}\PY{p}{;}
\end{Verbatim}

    \begin{center}
    \adjustimage{max size={0.9\linewidth}{0.9\paperheight}}{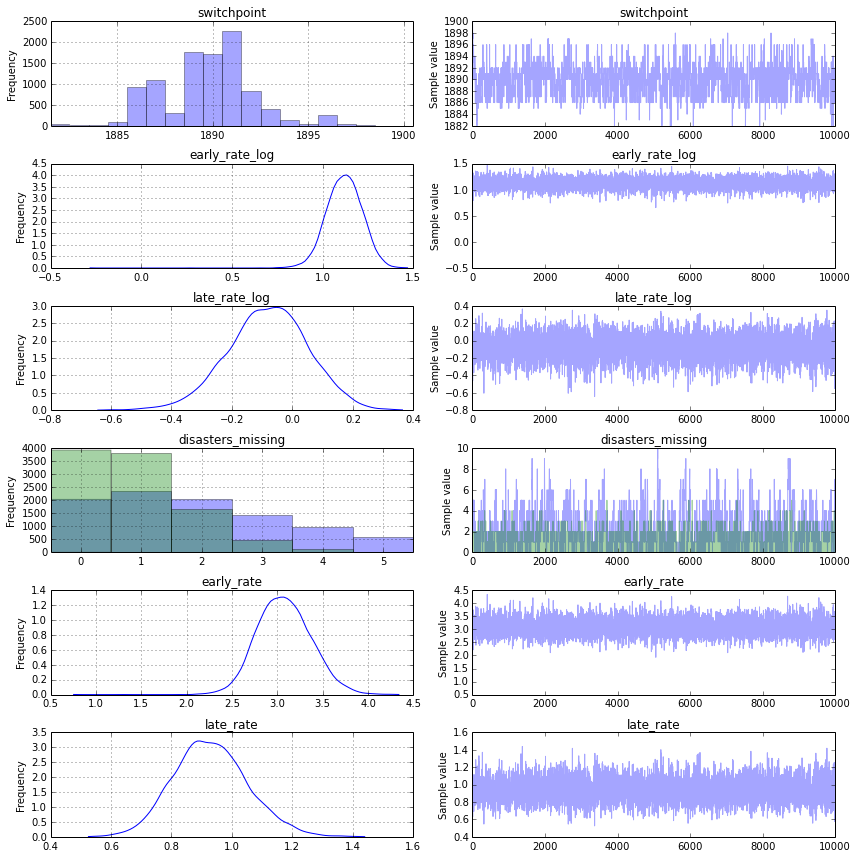}
    \end{center}
    { \hspace*{\fill} \\}

    \section{Arbitrary deterministics}\label{arbitrary-deterministics}

Due to its reliance on Theano, PyMC3 provides many mathematical
functions and operators for transforming random variables into new
random variables. However, the library of functions in Theano is not
exhaustive, therefore Theano and PyMC3 provide functionality for
creating arbitrary Theano functions in pure Python, and including these
functions in PyMC models. This is supported with the \texttt{as\_op}
function decorator.

Theano needs to know the types of the inputs and outputs of a function,
which are specified for \texttt{as\_op} by \texttt{itypes} for inputs
and \texttt{otypes} for outputs. The Theano documentation includes
\href{http://deeplearning.net/software/theano/library/tensor/basic.html\#all-fully-typed-constructors}{an
overview of the available types}.

    \begin{Verbatim}[commandchars=\\\{\}]
{\color{incolor}In [{\color{incolor}22}]:} \PY{k+kn}{import} \PY{n+nn}{theano.tensor} \PY{k+kn}{as} \PY{n+nn}{T}
         \PY{k+kn}{from} \PY{n+nn}{theano.compile.ops} \PY{k+kn}{import} \PY{n}{as\PYZus{}op}

         \PY{n+nd}{@as\PYZus{}op}\PY{p}{(}\PY{n}{itypes}\PY{o}{=}\PY{p}{[}\PY{n}{T}\PY{o}{.}\PY{n}{lscalar}\PY{p}{]}\PY{p}{,} \PY{n}{otypes}\PY{o}{=}\PY{p}{[}\PY{n}{T}\PY{o}{.}\PY{n}{lscalar}\PY{p}{]}\PY{p}{)}
         \PY{k}{def} \PY{n+nf}{crazy\PYZus{}modulo3}\PY{p}{(}\PY{n}{value}\PY{p}{)}\PY{p}{:}
             \PY{k}{if} \PY{n}{value} \PY{o}{\PYZgt{}} \PY{l+m+mi}{0}\PY{p}{:}
                 \PY{k}{return} \PY{n}{value} \PY{o}{\PYZpc{}} \PY{l+m+mi}{3}
             \PY{k}{else} \PY{p}{:}
                 \PY{k}{return} \PY{p}{(}\PY{o}{\PYZhy{}}\PY{n}{value} \PY{o}{+} \PY{l+m+mi}{1}\PY{p}{)} \PY{o}{\PYZpc{}} \PY{l+m+mi}{3}

         \PY{k}{with} \PY{n}{Model}\PY{p}{(}\PY{p}{)} \PY{k}{as} \PY{n}{model\PYZus{}deterministic}\PY{p}{:}
             \PY{n}{a} \PY{o}{=} \PY{n}{Poisson}\PY{p}{(}\PY{l+s}{\PYZsq{}}\PY{l+s}{a}\PY{l+s}{\PYZsq{}}\PY{p}{,} \PY{l+m+mi}{1}\PY{p}{)}
             \PY{n}{b} \PY{o}{=} \PY{n}{crazy\PYZus{}modulo3}\PY{p}{(}\PY{n}{a}\PY{p}{)}
\end{Verbatim}

    An important drawback of this approach is that it is not possible for
\texttt{theano} to inspect these functions in order to compute the
gradient required for the Hamiltonian-based samplers. Therefore, it is
not possible to use the HMC or NUTS samplers for a model that uses such
an operator. However, it is possible to add a gradient if we inherit
from \texttt{theano.Op} instead of using \texttt{as\_op}. The PyMC
example set includes
\href{https://github.com/pymc-devs/pymc/blob/master/pymc/examples/disaster_model_arbitrary_deterministic.py}{a
more elaborate example of the usage of \texttt{as\_op}}.

    \section{Arbitrary distributions}\label{arbitrary-distributions}

Similarly, the library of statistical distributions in PyMC3 is not
exhaustive, but PyMC allows for the creation of user-defined functions
for an arbitrary probability distribution. For simple statistical
distributions, the \texttt{DensityDist} function takes as an argument
any function that calculates a log-probability $log(p(x))$. This
function may employ other random variables in its calculation. Here is
an example inspired by a blog post by Jake Vanderplas on which priors to
use for a linear regression (Vanderplas, 2014).

\begin{Shaded}
\begin{Highlighting}[]
\CharTok{import} \NormalTok{theano.tensor }\CharTok{as} \NormalTok{T}
\CharTok{from} \NormalTok{pymc3 }\CharTok{import} \NormalTok{DensityDist, Uniform}

\KeywordTok{with} \NormalTok{Model() }\CharTok{as} \NormalTok{model:}
    \NormalTok{alpha = Uniform(}\StringTok{'intercept'}\NormalTok{, -}\DecValTok{100}\NormalTok{, }\DecValTok{100}\NormalTok{)}

    \CommentTok{# Create custom densities}
    \NormalTok{beta = DensityDist(}\StringTok{'beta'}\NormalTok{, }\KeywordTok{lambda} \NormalTok{value: -}\FloatTok{1.5} \NormalTok{* T.log(}\DecValTok{1} \NormalTok{+ value**}\DecValTok{2}\NormalTok{), testval=}\DecValTok{0}\NormalTok{)}
    \NormalTok{eps = DensityDist(}\StringTok{'eps'}\NormalTok{, }\KeywordTok{lambda} \NormalTok{value: -T.log(T.abs_(value)), testval=}\DecValTok{1}\NormalTok{)}

    \CommentTok{# Create likelihood}
    \NormalTok{like = Normal(}\StringTok{'y_est'}\NormalTok{, mu=alpha + beta * X, sd=eps, observed=Y)}
\end{Highlighting}
\end{Shaded}

    For more complex distributions, one can create a subclass of
\texttt{Continuous} or \texttt{Discrete} and provide the custom
\texttt{logp} function, as required. This is how the built-in
distributions in PyMC are specified. As an example, fields like
psychology and astrophysics have complex likelihood functions for a
particular process that may require numerical approximation. In these
cases, it is impossible to write the function in terms of predefined
theano operators and we must use a custom theano operator using
\texttt{as\_op} or inheriting from \texttt{theano.Op}.

Implementing the \texttt{beta} variable above as a \texttt{Continuous}
subclass is shown below, along with a sub-function using the
\texttt{as\_op} decorator, though this is not strictly necessary.

    \begin{Verbatim}[commandchars=\\\{\}]
{\color{incolor}In [{\color{incolor}23}]:} \PY{k+kn}{from} \PY{n+nn}{pymc3.distributions} \PY{k+kn}{import} \PY{n}{Continuous}

         \PY{k}{class} \PY{n+nc}{Beta}\PY{p}{(}\PY{n}{Continuous}\PY{p}{)}\PY{p}{:}
             \PY{k}{def} \PY{n+nf}{\PYZus{}\PYZus{}init\PYZus{}\PYZus{}}\PY{p}{(}\PY{n+nb+bp}{self}\PY{p}{,} \PY{n}{mu}\PY{p}{,} \PY{o}{*}\PY{n}{args}\PY{p}{,} \PY{o}{*}\PY{o}{*}\PY{n}{kwargs}\PY{p}{)}\PY{p}{:}
                 \PY{n+nb}{super}\PY{p}{(}\PY{n}{Beta}\PY{p}{,} \PY{n+nb+bp}{self}\PY{p}{)}\PY{o}{.}\PY{n}{\PYZus{}\PYZus{}init\PYZus{}\PYZus{}}\PY{p}{(}\PY{o}{*}\PY{n}{args}\PY{p}{,} \PY{o}{*}\PY{o}{*}\PY{n}{kwargs}\PY{p}{)}
                 \PY{n+nb+bp}{self}\PY{o}{.}\PY{n}{mu} \PY{o}{=} \PY{n}{mu}
                 \PY{n+nb+bp}{self}\PY{o}{.}\PY{n}{mode} \PY{o}{=} \PY{n}{mu}

             \PY{k}{def} \PY{n+nf}{logp}\PY{p}{(}\PY{n+nb+bp}{self}\PY{p}{,} \PY{n}{value}\PY{p}{)}\PY{p}{:}
                 \PY{n}{mu} \PY{o}{=} \PY{n+nb+bp}{self}\PY{o}{.}\PY{n}{mu}
                 \PY{k}{return} \PY{n}{beta\PYZus{}logp}\PY{p}{(}\PY{n}{value} \PY{o}{\PYZhy{}} \PY{n}{mu}\PY{p}{)}

         \PY{n+nd}{@as\PYZus{}op}\PY{p}{(}\PY{n}{itypes}\PY{o}{=}\PY{p}{[}\PY{n}{T}\PY{o}{.}\PY{n}{dscalar}\PY{p}{]}\PY{p}{,} \PY{n}{otypes}\PY{o}{=}\PY{p}{[}\PY{n}{T}\PY{o}{.}\PY{n}{dscalar}\PY{p}{]}\PY{p}{)}
         \PY{k}{def} \PY{n+nf}{beta\PYZus{}logp}\PY{p}{(}\PY{n}{value}\PY{p}{)}\PY{p}{:}
             \PY{k}{return} \PY{o}{\PYZhy{}}\PY{l+m+mf}{1.5} \PY{o}{*} \PY{n}{np}\PY{o}{.}\PY{n}{log}\PY{p}{(}\PY{l+m+mi}{1} \PY{o}{+} \PY{p}{(}\PY{n}{value}\PY{p}{)}\PY{o}{*}\PY{o}{*}\PY{l+m+mi}{2}\PY{p}{)}


         \PY{k}{with} \PY{n}{Model}\PY{p}{(}\PY{p}{)} \PY{k}{as} \PY{n}{model}\PY{p}{:}
             \PY{n}{beta} \PY{o}{=} \PY{n}{Beta}\PY{p}{(}\PY{l+s}{\PYZsq{}}\PY{l+s}{slope}\PY{l+s}{\PYZsq{}}\PY{p}{,} \PY{n}{mu}\PY{o}{=}\PY{l+m+mi}{0}\PY{p}{,} \PY{n}{testval}\PY{o}{=}\PY{l+m+mi}{0}\PY{p}{)}
\end{Verbatim}

    \section{Generalized Linear Models}\label{generalized-linear-models}

Generalized Linear Models (GLMs) are a class of flexible models that are
widely used to estimate regression relationships between a single
outcome variable and one or multiple predictors. Because these models
are so common, \texttt{PyMC3} offers a \texttt{glm} submodule that
allows flexible creation of various GLMs with an intuitive
\texttt{R}-like syntax that is implemented via the \texttt{patsy}
module.

The \texttt{glm} submodule requires data to be included as a
\texttt{pandas} \texttt{DataFrame}. Hence, for our linear regression
example:

    \begin{Verbatim}[commandchars=\\\{\}]
{\color{incolor}In [{\color{incolor}24}]:} \PY{c}{\PYZsh{} Convert X and Y to a pandas DataFrame}
         \PY{k+kn}{import} \PY{n+nn}{pandas}
         \PY{n}{df} \PY{o}{=} \PY{n}{pandas}\PY{o}{.}\PY{n}{DataFrame}\PY{p}{(}\PY{p}{\PYZob{}}\PY{l+s}{\PYZsq{}}\PY{l+s}{x1}\PY{l+s}{\PYZsq{}}\PY{p}{:} \PY{n}{X1}\PY{p}{,} \PY{l+s}{\PYZsq{}}\PY{l+s}{x2}\PY{l+s}{\PYZsq{}}\PY{p}{:} \PY{n}{X2}\PY{p}{,} \PY{l+s}{\PYZsq{}}\PY{l+s}{y}\PY{l+s}{\PYZsq{}}\PY{p}{:} \PY{n}{Y}\PY{p}{\PYZcb{}}\PY{p}{)}
\end{Verbatim}

    The model can then be very concisely specified in one line of code.

    \begin{Verbatim}[commandchars=\\\{\}]
{\color{incolor}In [{\color{incolor}25}]:} \PY{k+kn}{from} \PY{n+nn}{pymc3.glm} \PY{k+kn}{import} \PY{n}{glm}

         \PY{k}{with} \PY{n}{Model}\PY{p}{(}\PY{p}{)} \PY{k}{as} \PY{n}{model\PYZus{}glm}\PY{p}{:}
             \PY{n}{glm}\PY{p}{(}\PY{l+s}{\PYZsq{}}\PY{l+s}{y \PYZti{} x1 + x2}\PY{l+s}{\PYZsq{}}\PY{p}{,} \PY{n}{df}\PY{p}{)}
\end{Verbatim}

    The error distribution, if not specified via the \texttt{family}
argument, is assumed to be normal. In the case of logistic regression,
this can be modified by passing in a \texttt{Binomial} family object.

    \begin{Verbatim}[commandchars=\\\{\}]
{\color{incolor}In [{\color{incolor}26}]:} \PY{k+kn}{from} \PY{n+nn}{pymc3.glm.families} \PY{k+kn}{import} \PY{n}{Binomial}

         \PY{n}{df\PYZus{}logistic} \PY{o}{=} \PY{n}{pandas}\PY{o}{.}\PY{n}{DataFrame}\PY{p}{(}\PY{p}{\PYZob{}}\PY{l+s}{\PYZsq{}}\PY{l+s}{x1}\PY{l+s}{\PYZsq{}}\PY{p}{:} \PY{n}{X1}\PY{p}{,} \PY{l+s}{\PYZsq{}}\PY{l+s}{x2}\PY{l+s}{\PYZsq{}}\PY{p}{:} \PY{n}{X2}\PY{p}{,} \PY{l+s}{\PYZsq{}}\PY{l+s}{y}\PY{l+s}{\PYZsq{}}\PY{p}{:} \PY{n}{Y} \PY{o}{\PYZgt{}} \PY{l+m+mi}{0}\PY{p}{\PYZcb{}}\PY{p}{)}

         \PY{k}{with} \PY{n}{Model}\PY{p}{(}\PY{p}{)} \PY{k}{as} \PY{n}{model\PYZus{}glm\PYZus{}logistic}\PY{p}{:}
             \PY{n}{glm}\PY{p}{(}\PY{l+s}{\PYZsq{}}\PY{l+s}{y \PYZti{} x1 + x2}\PY{l+s}{\PYZsq{}}\PY{p}{,} \PY{n}{df\PYZus{}logistic}\PY{p}{,} \PY{n}{family}\PY{o}{=}\PY{n}{Binomial}\PY{p}{(}\PY{p}{)}\PY{p}{)}
\end{Verbatim}

    \section{Backends}\label{backends}

\texttt{PyMC3} has support for different ways to store samples during
and after sampling, called backends, including in-memory (default), text
file, and SQLite. These can be found in \texttt{pymc.backends}:

By default, an in-memory \texttt{ndarray} is used but if the samples
would get too large to be held in memory we could use the
\texttt{sqlite} backend:

    \begin{Verbatim}[commandchars=\\\{\}]
{\color{incolor}In [{\color{incolor}27}]:} \PY{k+kn}{from} \PY{n+nn}{pymc3.backends} \PY{k+kn}{import} \PY{n}{SQLite}

         \PY{k}{with} \PY{n}{model\PYZus{}glm\PYZus{}logistic}\PY{p}{:}
             \PY{n}{backend} \PY{o}{=} \PY{n}{SQLite}\PY{p}{(}\PY{l+s}{\PYZsq{}}\PY{l+s}{trace.sqlite}\PY{l+s}{\PYZsq{}}\PY{p}{)}
             \PY{n}{trace} \PY{o}{=} \PY{n}{sample}\PY{p}{(}\PY{l+m+mi}{5000}\PY{p}{,} \PY{n}{Metropolis}\PY{p}{(}\PY{p}{)}\PY{p}{,} \PY{n}{trace}\PY{o}{=}\PY{n}{backend}\PY{p}{)}
\end{Verbatim}

    \begin{Verbatim}[commandchars=\\\{\}]
[-----------------100\%-----------------] 5000 of 5000 complete in 1.7 sec
    \end{Verbatim}

    \begin{Verbatim}[commandchars=\\\{\}]
{\color{incolor}In [{\color{incolor}28}]:} \PY{n}{summary}\PY{p}{(}\PY{n}{trace}\PY{p}{,} \PY{n+nb}{vars}\PY{o}{=}\PY{p}{[}\PY{l+s}{\PYZsq{}}\PY{l+s}{x1}\PY{l+s}{\PYZsq{}}\PY{p}{,} \PY{l+s}{\PYZsq{}}\PY{l+s}{x2}\PY{l+s}{\PYZsq{}}\PY{p}{]}\PY{p}{)}
\end{Verbatim}

    \begin{Verbatim}[commandchars=\\\{\}]
x1:

  Mean             SD               MC Error         95\% HPD interval
  -------------------------------------------------------------------

  0.141            5.371            0.522            [-10.872, 11.579]

  Posterior quantiles:
  2.5            25             50             75             97.5
  |--------------|==============|==============|--------------|

  -11.702        -2.722         -0.043         2.959          11.026


x2:

  Mean             SD               MC Error         95\% HPD interval
  -------------------------------------------------------------------

  -0.627           26.921           2.620            [-56.304, 57.165]

  Posterior quantiles:
  2.5            25             50             75             97.5
  |--------------|==============|==============|--------------|

  -55.090        -14.627        0.022          13.840         59.016
    \end{Verbatim}

    The stored trace can then later be loaded using the \texttt{load}
command:

    \begin{Verbatim}[commandchars=\\\{\}]
{\color{incolor}In [{\color{incolor}29}]:} \PY{k+kn}{from} \PY{n+nn}{pymc3.backends.sqlite} \PY{k+kn}{import} \PY{n}{load}

         \PY{k}{with} \PY{n}{basic\PYZus{}model}\PY{p}{:}
             \PY{n}{trace\PYZus{}loaded} \PY{o}{=} \PY{n}{load}\PY{p}{(}\PY{l+s}{\PYZsq{}}\PY{l+s}{trace.sqlite}\PY{l+s}{\PYZsq{}}\PY{p}{)}
\end{Verbatim}

    More information about \texttt{backends} can be found in the docstring
of \texttt{pymc.backends}.

    \section{References}\label{references}

Patil, A., D. Huard and C.J. Fonnesbeck. (2010) PyMC: Bayesian
Stochastic Modelling in Python. Journal of Statistical Software, 35(4),
pp.~1-81

Bastien, F., Lamblin, P., Pascanu, R., Bergstra, J., Goodfellow, I.,
Bergeron, A., Bouchard, N., Warde-Farley, D., and Bengio, Y. (2012)
``Theano: new features and speed improvements''. NIPS 2012 deep learning
workshop.

Bergstra, J., Breuleux, O., Bastien, F., Lamblin, P., Pascanu, R.,
Desjardins, G., Turian, J., Warde-Farley, D., and Bengio, Y. (2010)
``Theano: A CPU and GPU Math Expression Compiler''. Proceedings of the
Python for Scientific Computing Conference (SciPy) 2010. June 30 - July
3, Austin, TX

Lunn, D.J., Thomas, A., Best, N., and Spiegelhalter, D. (2000) WinBUGS
-- a Bayesian modelling framework: concepts, structure, and
extensibility. Statistics and Computing, 10:325--337.

Neal, R.M. Slice sampling. Annals of Statistics. (2003).
doi:10.2307/3448413.

van Rossum, G. The Python Library Reference Release 2.6.5., (2010). URL
http://docs.python.org/library/.

Duane, S., Kennedy, A. D., Pendleton, B. J., and Roweth, D. (1987)
``Hybrid Monte Carlo'', Physics Letters, vol.~195, pp.~216-222.

Stan Development Team. (2014). Stan: A C++ Library for Probability and
Sampling, Version 2.5.0. http://mc-stan.org.

Gamerman, D. Markov Chain Monte Carlo: statistical simulation for
Bayesian inference. Chapman and Hall, 1997.

Hoffman, M. D., \& Gelman, A. (2014). The No-U-Turn Sampler: Adaptively
Setting Path Lengths in Hamiltonian Monte Carlo. The Journal of Machine
Learning Research, 30.

Vanderplas, Jake. ``Frequentism and Bayesianism IV: How to be a Bayesian
in Python.'' Pythonic Perambulations. N.p., 14 Jun 2014. Web. 27 May.
2015.
\url{https://jakevdp.github.io/blog/2014/06/14/frequentism-and-bayesianism-4-bayesian-in-python/}.

R.G. Jarrett. A note on the intervals between coal mining disasters.
Biometrika, 66:191--193, 1979.


    \end{document}